\documentclass[superscriptaddress,twocolumn,showpacs,prb]{revtex4-1}
\usepackage[utf8]{inputenc} 
\usepackage{mathtools}
\usepackage{amsmath}
\usepackage{braket}
\usepackage{comment}
\usepackage{booktabs} 
\usepackage{siunitx} 
\usepackage{booktabs}
\usepackage{graphicx}
\usepackage{multirow}
\usepackage{graphicx}
\usepackage{pgfplots}
\pgfplotsset{compat=1.14}
\usepackage{csquotes}
\usepackage{hhline}
\usepackage{tabularx}
\usepackage{subcaption}
\usepackage{amssymb}
\usepackage{amsmath}
\usepackage{braket}
\usepackage{graphicx}
\usepackage{tikz}
\usepackage{geometry}
\usepackage{quantikz}
\usepackage{ulem} 
\usepackage{textcomp} 
\usepackage{dcolumn}
\usepackage{tabularx}
\usepackage{braket}
\usepackage{float}
\usepackage{listings}
\usepackage{color}
\usepackage{subcaption}
\usepackage{placeins}
\definecolor{mygreen}{rgb}{0,0.6,0}
\definecolor{mygray}{rgb}{0.5,0.5,0.5}
\definecolor{mymauve}{rgb}{0.58,0,0.82}

\newcolumntype{C}{>{\centering\arraybackslash}X}

\begin{document}

\title{Excitation Gaps in the Fermi-Hubbard Model via Variational Quantum Eigensolver}
\author{Mrinal Dev}
\email{mrinaldev3@gmail.com}
\affiliation{National Institute of Technology Rourkela 769008, Odisha, India}
\author{Bikash K. Behera}
\email{bikas.riki@gmail.com}
\affiliation{Bikash's Quantum (OPC) Private Limited, Mohanpur 741246, West Bengal, India}
\author{Vivek Vyas}
\affiliation{Indian Institute of Information Technology Vadodara Gandhinagar, Gujarat, India}
\author{Prasanta K. Panigrahi}
\email{pprasanta@iiserkol.ac.in}
\affiliation{Department of Physical Sciences,\\ Indian Institute of Science Education and Research Kolkata, Mohanpur 741246, West Bengal, India}
\affiliation{CQST, Siksha O Anusandhan, Khandagiri, Bhubaneswar, 751030, Odisha, India}

\begin{abstract}
The Hubbard model is a challenging quantum many-body problem and serves as a benchmark for quantum computing research. Accurate computation of its ground and excited state energies is essential for understanding correlated electron systems. In this study, the ground, first, and second excited state energies of 4$\times$1 and 2$\times$2 Hubbard lattices are obtained using a newly designed ansatz circuit. 
The ansatz is constructed by combining concepts from the Hamiltonian Variational Ansatz (HVA) and the Number-Preserving Ansatz (NPA). A hybrid optimization strategy is applied, where COBYLA is used for coarse convergence and L-BFGS for fine-tuning. The resulting energies are evaluated, and the corresponding physical properties of the systems are analyzed through phase diagrams of the energy excitation gaps for different charge and spin configurations.
\end{abstract}
 

\maketitle

\section{Introduction}

The Fermi-Hubbard model is a theoretical framework in condensed matter physics that describes the behavior of interacting electrons on a lattice system, where each lattice site can host at most two electrons with opposite spins due to the Pauli exclusion principle \cite{leblanc2015solutions}. 
Electrons in the model face a competition between kinetic energy, which promotes hopping to neighboring sites, and on-site Coulomb repulsion, which penalizes double occupancy. 
This competition generates a rich landscape of strongly correlated electron phenomena \cite{dagotto2005complexity,fradkin2015colloquium,PhysRevB.66.075128}. Every system tries to minimize its energy to its ground state. The electrons in the Fermi-Hubbard model also try to arrange themselves in a way that they reach the ground state. This ground state, although it is not very trivial and depends on the configuration and number of electrons, requires high computational strength to find.
This work not only focuses on the ground state but also the excited states of the Fermi Hubbard model and its physical implications.
Understanding such many-electron systems is critical because the Hubbard model captures a variety of emergent properties observed in real materials. These include high-temperature superconductivity \cite{capone2002strongly,li2019superconductivity}, Mott insulating behavior \cite{grytsiuk2024nb3cl8}, non-Fermi liquid phases \cite{stewart2001non,kotliar1993quantum}, excitonic effects \cite{PhysRevB.85.165135}, and non-trivial magnetism \cite{PhysRevB.93.054429,guo2018antiferromagnetic,PhysRevLett.62.591}. Superconductivity in the Hubbard model can emerge when electrons of opposite spin form Cooper pairs on lattice sites with attractive interactions, which can occur by effectively switching the sign of the on-site interaction \cite{jiang2018superconductivity}. The relationship between pseudogap behavior and superconductivity has also been explored within this framework \cite{gull2013superconductivity}. Furthermore, the model has been used to investigate Mott transitions \cite{kancharla2007band,lee2014competition} and the interplay of competing and intertwined electronic orders \cite{dagotto2005complexity,fradkin2015colloquium,PhysRevB.66.075128}.  

The dimensionality of the lattice plays a central role in the solvability of the Hubbard model. The one-dimensional (1D) model admits an exact analytical solution through the Bethe ansatz, whereas the two-dimensional (2D) case remains unsolved analytically, making computational studies indispensable \cite{leblanc2015solutions}. Classical numerical methods such as exact diagonalization, Density Matrix Renormalization Group (DMRG), and Quantum Monte Carlo (QMC) have been widely applied to study the Fermi-Hubbard model \cite{leblanc2015solutions}. Exact diagonalization provides accurate results but is limited by the exponential growth of the Hilbert space with system size. DMRG is efficient in 1D but less effective in higher dimensions, and QMC methods face severe limitations due to the fermionic sign problem, which leads to slow convergence and large uncertainties in certain regimes \cite{leblanc2015solutions}. Even with state-of-the-art supercomputing resources, exact diagonalization has been restricted to systems of approximately 17 electrons on 22 lattice sites \cite{yamada200516}.  

Given the exponential complexity of classical methods, quantum computing has emerged as a promising alternative. The Variational Quantum Eigensolver (VQE), first demonstrated by Peruzzo \textit{et al.} \cite{peruzzo2014variational} and formalized by McClean \textit{et al.} \cite{mcclean2016theory}, provides a hybrid quantum-classical framework for finding the ground and excited states of quantum systems. In VQE, a quantum processor prepares a parameterized ansatz (guess) state, and the energy expectation value is measured. A classical optimizer then iteratively updates the parameters to minimize the energy. The hybrid structure makes VQE well-suited for noisy intermediate-scale quantum (NISQ) devices, which currently have limited qubit counts and coherence times \cite{deglmann2015application}. The choice of ansatz is crucial because a well-designed ansatz reduces circuit depth and measurement error. Poor ansatz selection can require deeper circuits and slow convergence, while effective ansatz choices can capture relevant physics with minimal overhead \cite{alvertis2025classical}. VQE has been applied to various fields, including quantum chemistry \cite{deglmann2015application}, drug discovery \cite{cao2018potential}, materials science \cite{lordi2021advances}, and chemical engineering \cite{cao2019quantum}. In the context of the Hubbard model, VQE has successfully computed ground state energies for lattice sizes up to 12 sites, demonstrating its utility for strongly correlated systems \cite{alvertis2025classical}.  

In this work, a modified ansatz is introduced for the Fermi-Hubbard model, combining elements of the Hamiltonian Variational Ansatz (HVA) and the Number-Preserving Ansatz (NPA) to capture low-energy states effectively. A hybrid optimization strategy is employed, using COBYLA for coarse convergence followed by L-BFGS for fine-tuning. This approach is applied to 4×1 and 2×2 Hubbard lattices to compute the ground, first, and second excited state energies. Physical properties such as spin and charge excitations are analyzed to provide further insight into the model behavior.  

The remainder of this paper is organized as follows: Section~\ref{background} presents the mathematical formulation of the Fermi-Hubbard model and a summary of relevant ansatz techniques for VQE. Section~\ref{methodology} introduces the proposed ansatz construction and hybrid optimization strategy. Section~\ref{results} presents the computed ground and excited state energies and the analysis of their physical properties. Finally, Section~\ref{conclusion} summarizes the results and discusses potential avenues for future research.

\section{Background}\label{background}

\begin{figure}
    \centering
    \includegraphics[width=\linewidth]{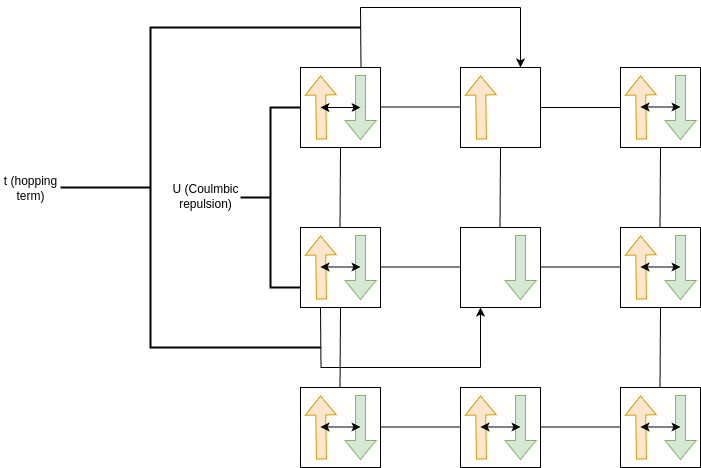}
    \caption{Structure of the 3$\times$3 two-dimensional Hubbard model.}
    \label{fig:Hubbard}
\end{figure}

The Fermi-Hubbard model consists of two fundamental components: (1) a kinetic energy term, which allows electrons to hop between lattice sites (in this case, restricted to nearest-neighbor hopping), and (2) a potential energy term arising from the Coulombic repulsion experienced when two electrons occupy the same site. The structural representation of the 2D $3 \times 3$ Hubbard model is illustrated in Fig.~\ref{fig:Hubbard}. Mathematically, the Fermi-Hubbard Hamiltonian can be expressed using the annihilation, creation, and number operators, as shown in Eq.~\ref{eq:hubbard_hamiltonian} \cite{altland2010condensed}. It contains two terms: a kinetic term that allows hopping between neighboring sites, and an onsite potential term that accounts for the repulsion due to double occupancy. The Hamiltonian is given by:

\begin{equation}
\label{eq:hubbard_hamiltonian}
\mathcal{H} = - \sum_{\langle i, j \rangle, \sigma} \left( a^{\dagger}_{i\sigma} a_{j\sigma} + a^{\dagger}_{j\sigma} a_{i\sigma} \right)
+ {U}\sum_{i} n_{i\uparrow} n_{i\downarrow},
\end{equation}

where $a^{\dagger}$ is the creation operator, $a$ is the annihilation operator, $n$ is the number operator, and $U$ represents the strength of the onsite Coulombic potential. To determine the ground-state energy, one evaluates the expectation value of the Hamiltonian over a parameterized ansatz (guess) state. This is implemented using a hybrid quantum-classical approach: the quantum processor evaluates the energy expectation value for a given set of parameters, which are then optimized iteratively by a classical processor. The classical processor updates the parameters and sends them back to the quantum processor, forming a feedback loop until the energy is minimized. As previously discussed, selecting an effective ansatz is crucial for the VQE. The term \textit{ansatz} literally means ``a guess," and here it refers to an initial guess for the wavefunction. A good ansatz can reach the ground state with lower circuit depth and reduced error, making its choice a critical design decision. A commonly employed ansatz for the Hubbard model is the HVA, inspired by the Quantum Approximate Optimization Algorithm (QAOA) \cite{farhi2014quantum} and the adiabatic theorem \cite{farhi2000quantum}. The adiabatic theorem states that if one starts with a state $H_a$, it can be evolved to state $H_b$,by applying a sequence of evolutions of the form $e^-{itH_A}$, $ e^ -{itH_B}$ for sufficiently small t. 
QAOA uses this concept by starting with a  simpler mixer Hamiltonian whose ground state can be easily prepared, and evolves it iteratively with a cost Hamiltonian whose ground state is to be found.
Unlike QAOA, which applies only two operators per layer, one for the cost function and another as a mixer, HVA is divided into non-commuting parts and applies a sequence of of these non-commuting terms of the Hamiltonian. Using this method Hamiltonian can be decomposed as follows \cite{cade2020strategies}:

\begin{equation}
H = \sum_{s} H_s,
\label{eq:hamiltonian_decomposition}
\end{equation}

where it is assumed that each pair of terms $H_s$ and $H_{s'}$ does not commute, i.e.,
\begin{eqnarray}
[H_s, H_{s'}] \neq 0
\end{eqnarray}
This process of evolution can be repeated over a fixed number of layers.
The corresponding variational state after $p$ layers is then given by:
\begin{equation}
|\psi_p\rangle = \prod_{\ell=1}^{p} \left( \prod_{s} \exp(-i \theta_{s,\ell} H_s) \right) |\psi_0\rangle
\label{eq:hamiltonian_ansatz}
\end{equation}

Another ansatz employed for the Fermi-Hubbard model is the NPA. This approach replaces all the circuit gates in HVA with a generalized two-qubit unitary operator defined as:

\begin{equation}
U_{\text{NP}}(\theta, \phi) =
\begin{pmatrix}
1 & 0 & 0 & 0 \\
0 & \cos\theta & i\sin\theta & 0 \\
0 & i\sin\theta & \cos\theta & 0 \\
0 & 0 & 0 & e^{i\phi}
\end{pmatrix}.
\label{eq:unp_gate}
\end{equation}

This formalism ensures number preservation in the quantum circuit while maintaining flexibility for variational optimization. It uses the same circuit structure as HVA, just replace each gate with the aforementioned unitary gate.
Using this ansatz, one can evolve to the ground state energy, but to achieve an excited state,a penalty term that imposes an orthogonality condition needs to be added, ensuring that when this new function optimizes, it does not optimize to the ground state, as overlap with the ground state will increase the energy value. There have been previous attempts at finding lower excited states of the Fermi-Hubbard model through this methodology \cite{zhang2025unified}
\section{Methodology}\label{methodology}
\begin{figure*}
\includegraphics[width=\linewidth]{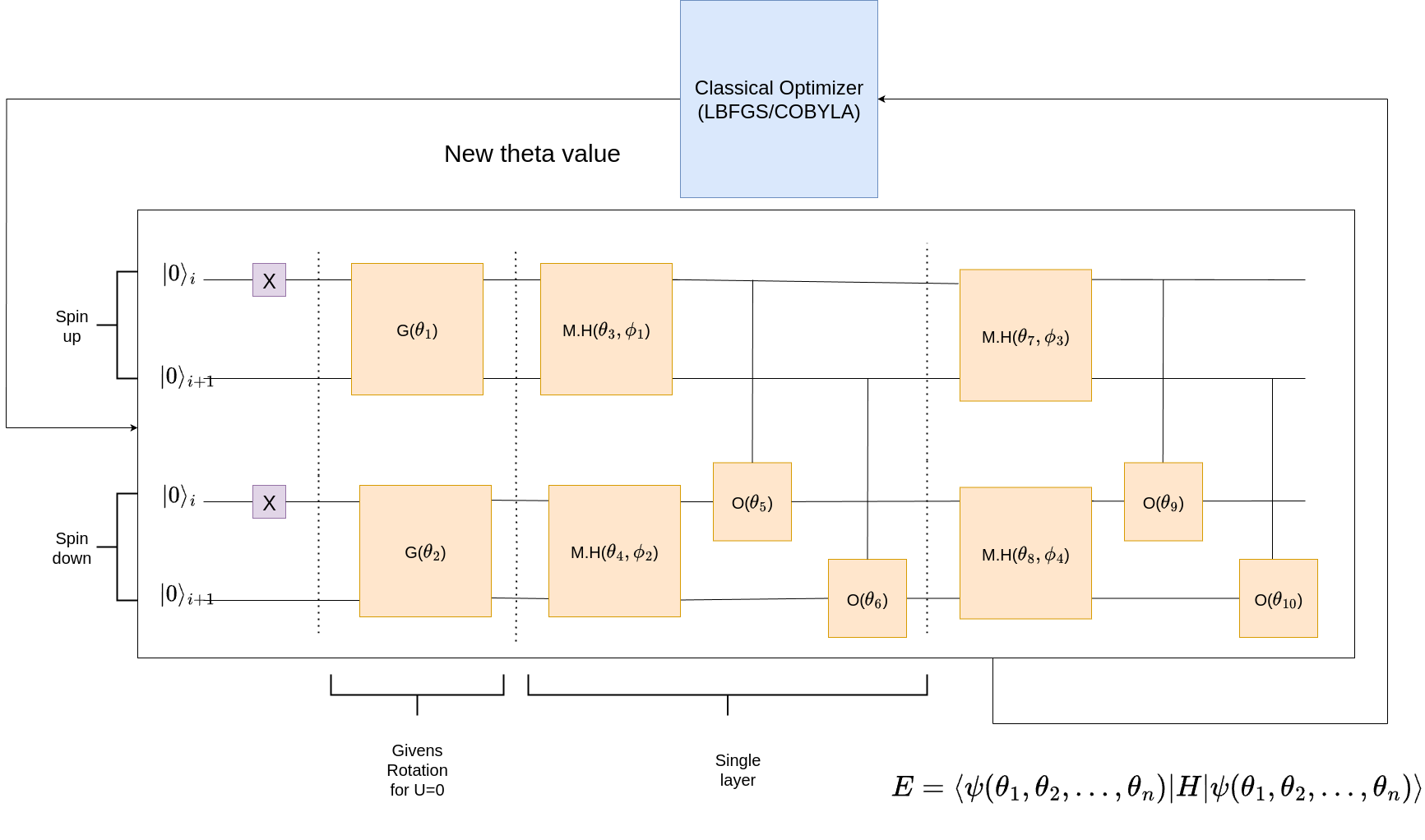}
\caption{In the circuit for the 2$\times$1 Hubbard model, the first two qubits are for spin up at adjacent lattice sites, followed by spin down on the same lattice site. We then apply G Givens' rotation, followed by M.H.'s modified hopping and onsite interaction.}
\label{fig:full_circuit}
\end{figure*}

Conventionally, HVA has been used as an ansatz to find the ground energy of the Fermi-Hubbard model cite.
An efficient HVA can also be constructed using fermionic swaps, which we will elaborate on later in the methodology \cite{stanisic2022observing}. In this work, we propose a new technique for ansatz construction that combines elements of both HVA and NPA. Unlike the NPA, where each gate is replaced by a generalized number-preserving unitary, here the ansatz retains the same gates as HVA. However, similar to NPA principles, each gate is assigned a unique variational parameter, and the structure of the HVA circuit is preserved;  the number of gates and the position and assignments of the gates are the same. Its circuit diagram is the same as HVA, but with each gate using a different parameter Fig.  \ref{fig:full_circuit}.
At first glance, using separate parameters for commuting terms might seem counterintuitive, as commuting operators share common eigenvalues and could, in principle, share parameters. Nevertheless, an ansatz in VQE is fundamentally a \textit{guess} and does not have to strictly adhere to these heuristics. By assigning independent parameters to all gates, even for commuting terms, we achieve better variational flexibility and faster convergence. This strategy allows the circuit to reach the ground state with a low circuit depth of just two layers \cite{stanisic2022observing}. The initial state is prepared using Pauli X operators, where the zero state is the absence of an electron and  1 state is the presence of an electron.
First, the zero potential state  is prepared using a given rotation \cite{jiang2018quantum}, represented by the matrix:

\begin{equation}\label{eq:givens_matrix}
G(\theta) = 
\begin{bmatrix}
1 & 0 & 0 & 0 \\
0 & \cos\theta/2 & -\sin\theta/2 & 0 \\
0 & \sin\theta/2 & \cos\theta/2 & 0 \\
0 & 0 & 0 & 1
\end{bmatrix}.
\end{equation}

\begin{figure*}[!t]
\centering
\begin{minipage}{0.5\textwidth}
\centering
$G(\theta) \equiv$ \quad
\begin{quantikz}
\lstick{$q_0$} & \ctrl{1} & \gate{R_y(\theta)} & \ctrl{1} & \qw \\
\lstick{$q_1$} & \targ{} & \ctrl{-1} & \targ{} & \qw
\end{quantikz}
\caption*{(a)}
\end{minipage}%
\hfill
\begin{minipage}{0.48\textwidth}
\centering
$H(\theta) \equiv$ \quad
\begin{quantikz}
\lstick{$q_0$} & \ctrl{1} & \gate{R_x(\theta)} & \ctrl{1} & \qw \\
\lstick{$q_1$} & \targ{} & \ctrl{-1} & \targ{} & \qw
\end{quantikz}
\caption*{(b)}
\end{minipage}

\vspace{0.7cm}
\begin{minipage}{0.5\textwidth}

$M.H(\theta,\phi) \equiv$ \quad
\begin{quantikz}
\lstick{$q_0$} & \ctrl{1} & \gate{R_z(\phi)} & \gate{R_x(\theta)} & \ctrl{1} & \qw \\
\lstick{$q_1$} & \targ{} & \ctrl{-1} & \ctrl{-1} & \targ{} & \qw
\end{quantikz}
\caption*{(c)}
\end{minipage}%
\hfill
\begin{minipage}{0.5\textwidth}
\centering
$O(\theta) \equiv$ \quad
\begin{quantikz}
\lstick{$q_0$} & \ctrl{1} & \qw \\
\lstick{$q_1$} & \gate{P(\theta)} & \qw
\end{quantikz}
\caption*{(d)}
\end{minipage}

\caption{Circuit definitions for the unitary operators. The operators (a) $G(\theta)$, (b) $H(\theta,\phi)$, (c) $M.H(\theta,\phi)$, and (d) $O(\theta)$ are defined by their corresponding quantum circuit representations.}
\label{fig:quantum_operators_equations}
\end{figure*}

The corresponding quantum circuit implementations are shown in Fig.~\ref{fig:quantum_operators_equations}. This initial circuit enables the system to reach the ground state of the non-interacting ($U = 0$) model. To extend this to the complete Hubbard Hamiltonian, we need to construct a Hamiltonian-inspired ansatz like HVA, which requires implementing the hopping term and the onsite term in (Eq.~\ref{eq:hubbard_hamiltonian}). This equation is in the form of creation and annihilation operators. To construct an ansatz from it, we need to transform it into terms of Pauli matrices, because that is the language of quantum circuits.
We apply the Jordan-Wigner transformation to achieve the form we require for both terms. After doing the transform, we can convert it into the matrix form since we know the matrix representation in terms of Pauli matrices, and then these matrices can now be implemented using any basic gates one wants, or are provided by the hardware being used.

\subsection{Hopping Term}

Applying the Jordan-Wigner transformation to the hopping term yields:

\begin{equation}\label{eq:hopping_transform}
a_i^\dagger a_j + a_j^\dagger a_i \rightarrow \frac{1}{2}(X_i X_j + Y_i Y_j) Z_{i+1} Z_{i+2} \ldots Z_{j-1}.
\end{equation}

\begin{figure}
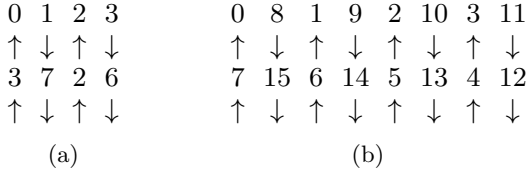

  \centering
  \begin{subfigure}[b]{0.49\linewidth}
    \centering
    \scalebox{1.15}{$
      \begin{array}{cccc}
      0 & 1 & 2 & 3 \\
      \uparrow & \downarrow & \uparrow & \downarrow \\
      3 & 7 & 2 & 6 \\
      \uparrow & \downarrow & \uparrow & \downarrow \\
      \end{array}
    $}
    \caption{}
    \label{fig:lattice_2x2}
  \end{subfigure}
  \hfill
  \begin{subfigure}[b]{0.49\linewidth}
    \centering
    \scalebox{1.15}{$
      \begin{array}{cccccccc}
      0 & 8 & 1 & 9 & 2 & 10 & 3 & 11 \\
      \uparrow & \downarrow & \uparrow & \downarrow & \uparrow & \downarrow & \uparrow & \downarrow \\
      7 & 15 & 6 & 14 & 5 & 13 & 4 & 12 \\
      \uparrow & \downarrow & \uparrow & \downarrow & \uparrow & \downarrow & \uparrow & \downarrow \\
      \end{array}
    $}
    \caption{}
    \label{fig:lattice_4x2}
  \end{subfigure}
  \caption{Lattice configurations with snake ordering: (a) 2×2 lattice and (b) 4×2 lattice.}
  \label{fig:lattice_all}
\end{figure}

Here, the lattice is indexed using \textit{snake ordering}, as illustrated in Figs.~\ref{fig:lattice_2x2} and \ref{fig:lattice_4x2}. Spin-up qubits are labeled first, followed by spin-down qubits, to maximize adjacency for same-spin hopping, which is necessary since only electrons with the same spin can hop without violating the Pauli exclusion principle. The transformation produces a string of $Z$-operators between non-adjacent sites. These $Z$-strings are absent for horizontal hopping but appear for vertical hopping. Fermionic SWAP gates are employed to remove them. These gates swap two qubits and introduce a negative sign to account for fermionic antisymmetry:

\begin{eqnarray}
\begin{bmatrix}
1 & 0 & 0 & 0 \\
0 & 0 & 1 & 0 \\
0 & 1 & 0 & 0 \\
0 & 0 & 0 & -1
\end{bmatrix}.
\end{eqnarray}

Hopping between adjacent spins is implemented as:

\begin{equation}
\label{eq:hopping_exponential}
e^{i\theta \left(\frac{XX + YY}{2}\right)},
\end{equation}

With the corresponding matrix representation:

\begin{equation}\label{eq:hopping_matrix}
H(\theta) = 
\begin{bmatrix}
0 & 0 & 0 & 0 \\
0 & \cos{\theta/2} & -i\sin{\theta/2} & 0 \\
0 & i\sin{\theta/2} & \cos{\theta/2} & 0 \\
0 & 0 & 0 & 0
\end{bmatrix}.
\end{equation}
\begin{equation}\label{eq:Mhopping_matrix}
M.H(\theta,\phi)=
\begin{pmatrix}
1 & 0 & 0 & 0 \\
0 & e^{i\phi/2} \cos(\theta/2) & i e^{i\phi/2} \sin(\theta/2) & 0 \\
0 & -i \sin(\theta/2) e^{-i\phi/2} & e^{-i\phi/2} \cos(\theta/2) & 0 \\
0 & 0 & 0 & 1
\end{pmatrix}
\end{equation}
The circuit has been enhanced to improve expressivity by including a controlled-$R_z$ (CRZ) gate, which adds a phase component to the variational evolution. This is again allowed from the initial argument that the ansatz does not need to adhere strictly to any rules. The corresponding circuit implementation is given in the \ref{fig:quantum_operators_equations}.

\subsection{On-Site Interaction}

The on-site interaction term transforms simply as:

\begin{equation}\label{eq:onsite_transform}
n_i n_j = c_i^\dagger c_i\, c_j^\dagger c_j \Rightarrow \ket{1}\bra{1} \otimes \ket{1}\bra{1}.
\end{equation}

Its variational evolution is expressed as:

\begin{equation}\label{eq:onsite_exponential}
e^{i\theta \ket{11}\bra{11}},
\end{equation}

with the corresponding matrix

\begin{figure}[H]
\centering
\begin{equation}\label{eq:onsite_matrix}
O(\theta)=
\begin{bmatrix}
1 & 0 & 0 & 0 \\
0 & 1 & 0 & 0 \\
0 & 0 & 1 & 0 \\
0 & 0 & 0 & e^{i\theta}
\end{bmatrix}
\end{equation}
\end{figure}

\subsection{Optimizer}

The VQE requires an optimizer to iteratively tune circuit parameters toward the ground state. Traditionally, a single classical optimizer is used; however, we adopt a hybrid optimization strategy combining COBYLA and L-BFGS-B. Specifically, COBYLA is employed for the first 500 iterations to perform global exploration, followed by 50 iterations of L-BFGS-B for rapid local convergence.

\section{Results \label{results}}

\begin{figure*}[t]
    \centering

    \begin{subfigure}[t]{0.32\textwidth}
        \includegraphics[width=\linewidth]{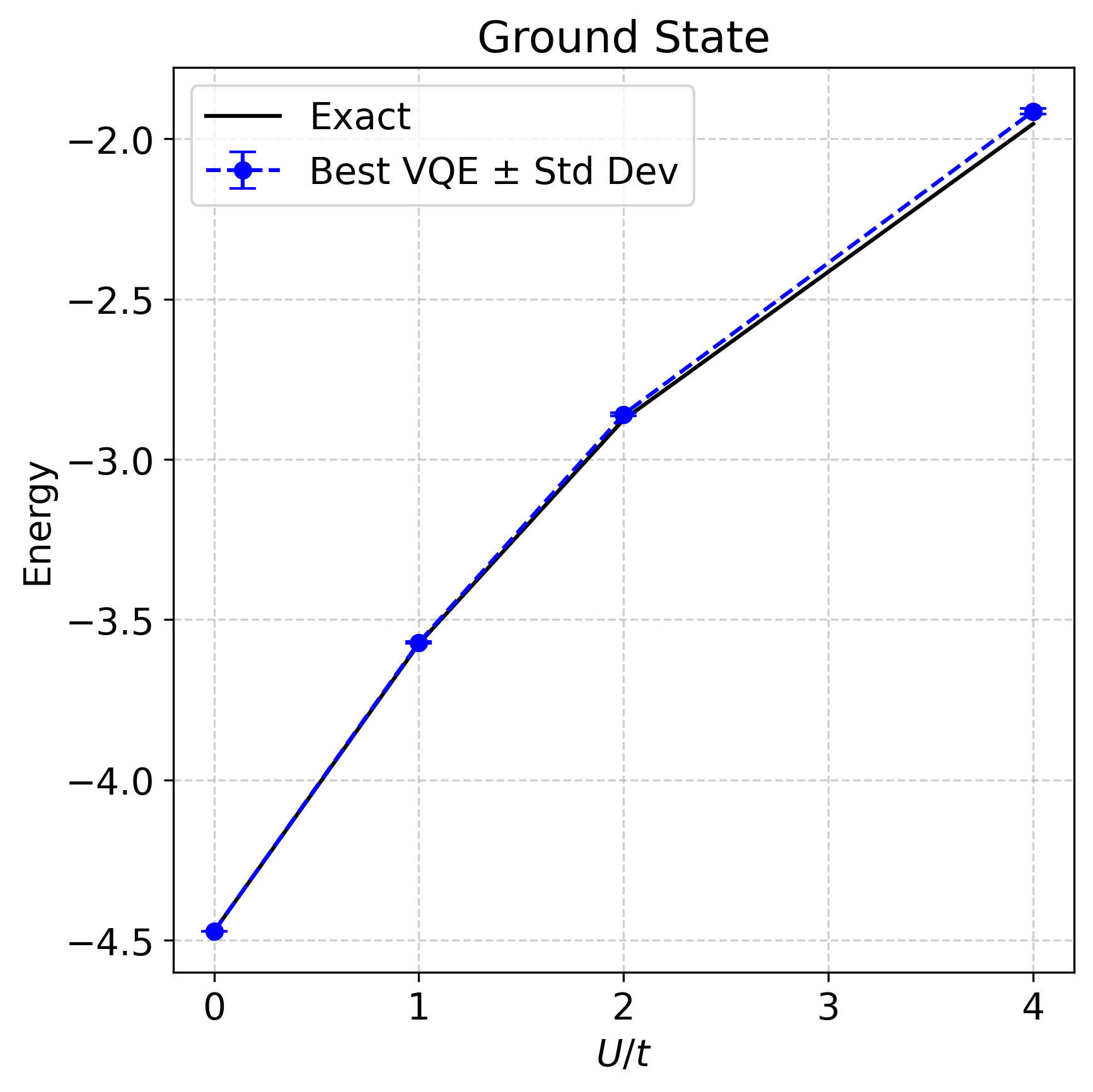}
        \caption{}

    \end{subfigure}
    \hfill
    \begin{subfigure}[t]{0.32\textwidth}
        \includegraphics[width=\linewidth]{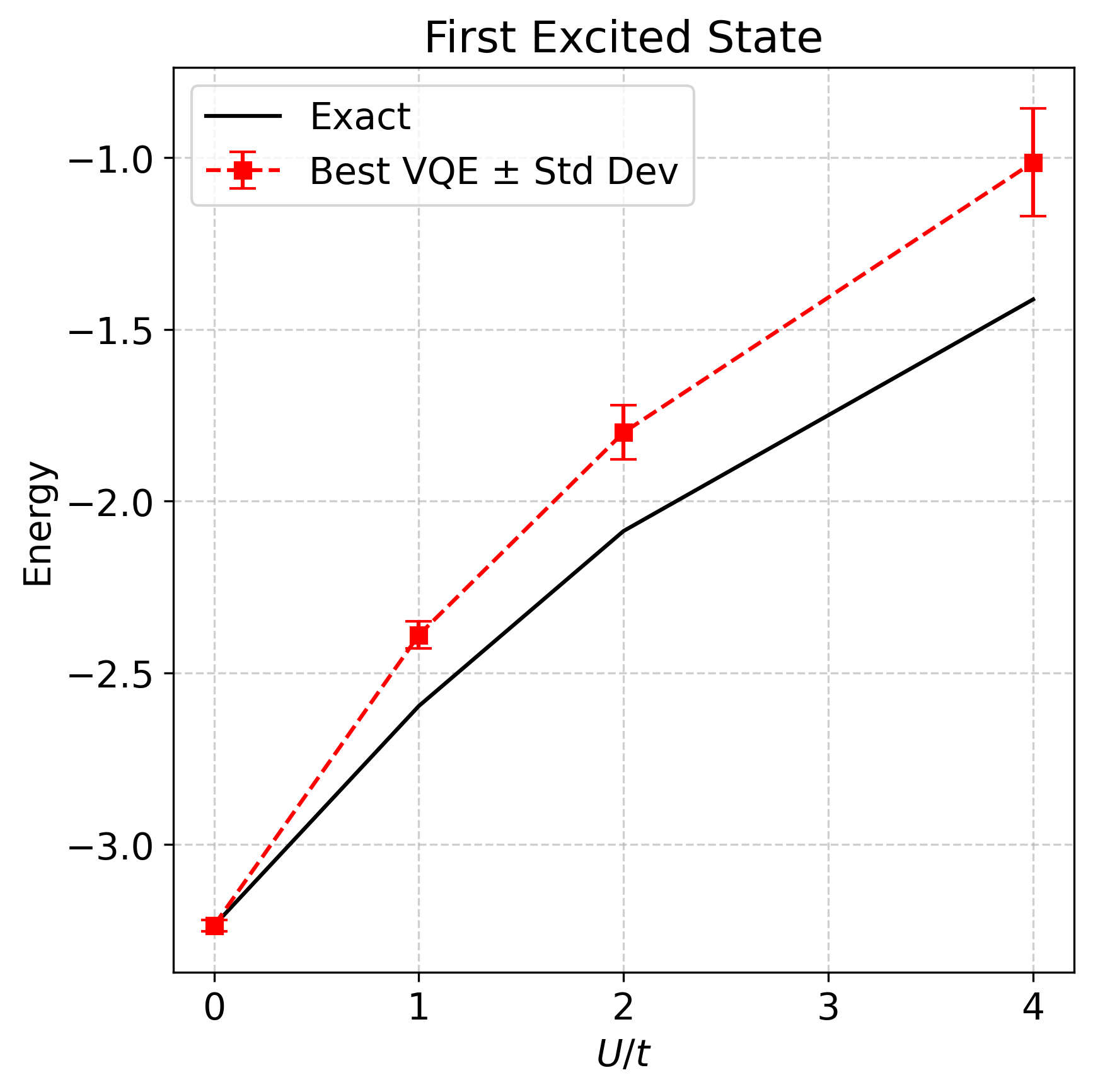}
        \caption{}
    \end{subfigure}
    \hfill
    \begin{subfigure}[t]{0.32\textwidth}
        \includegraphics[width=\linewidth]{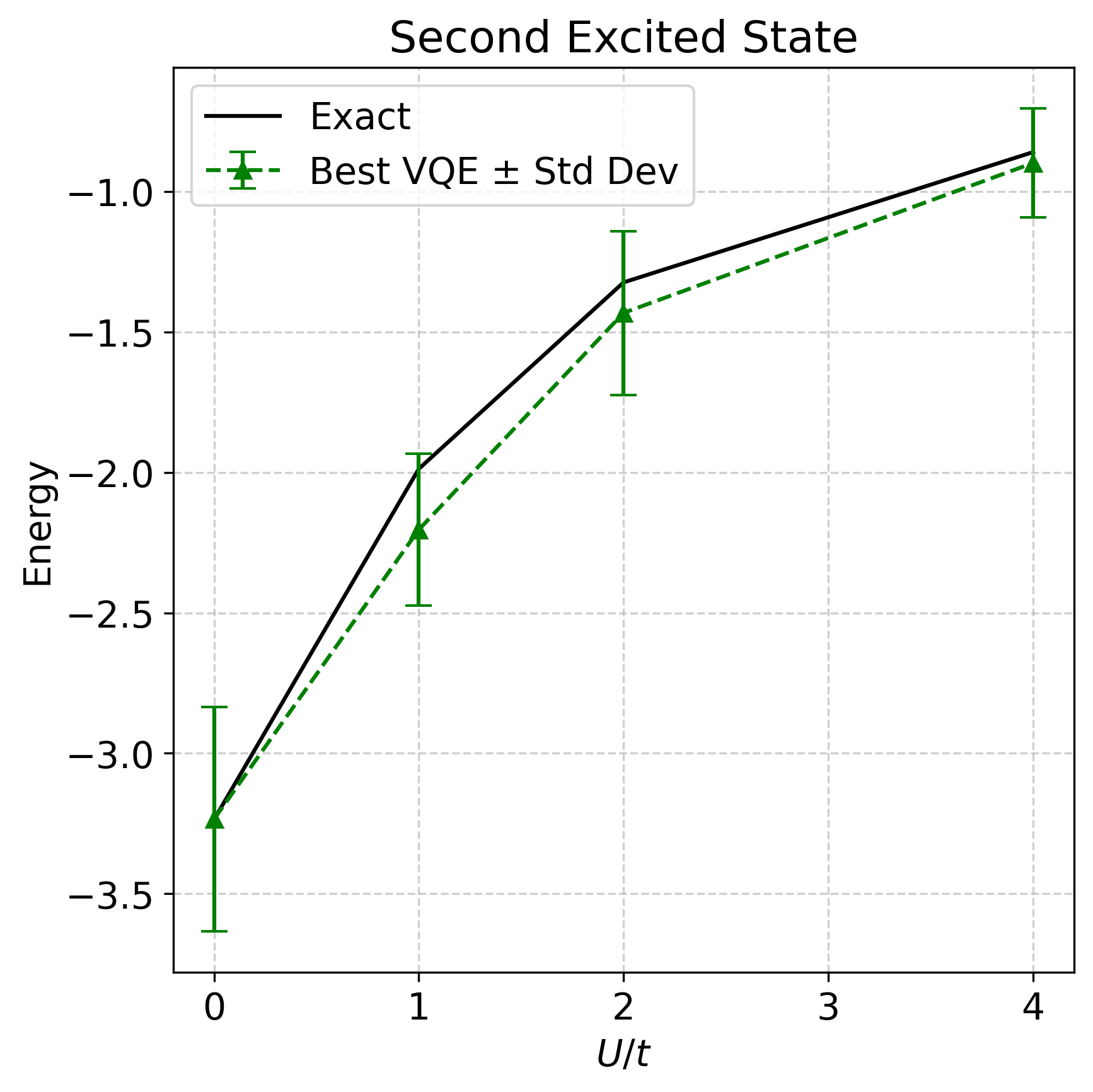}
        \caption{}
    \end{subfigure}


    \caption{(a) Ground, (b) first excited state, (c) second excited energy plotted with exact diagonalization result and VQE, with the horizontal axis U/t =0,1,2,4, for 4$\times$1.}
    
    \label{fig:4*1}
\end{figure*}

\begin{figure*}[t]
    \centering

    \begin{subfigure}[t]{0.32\textwidth}
        \includegraphics[width=\linewidth]{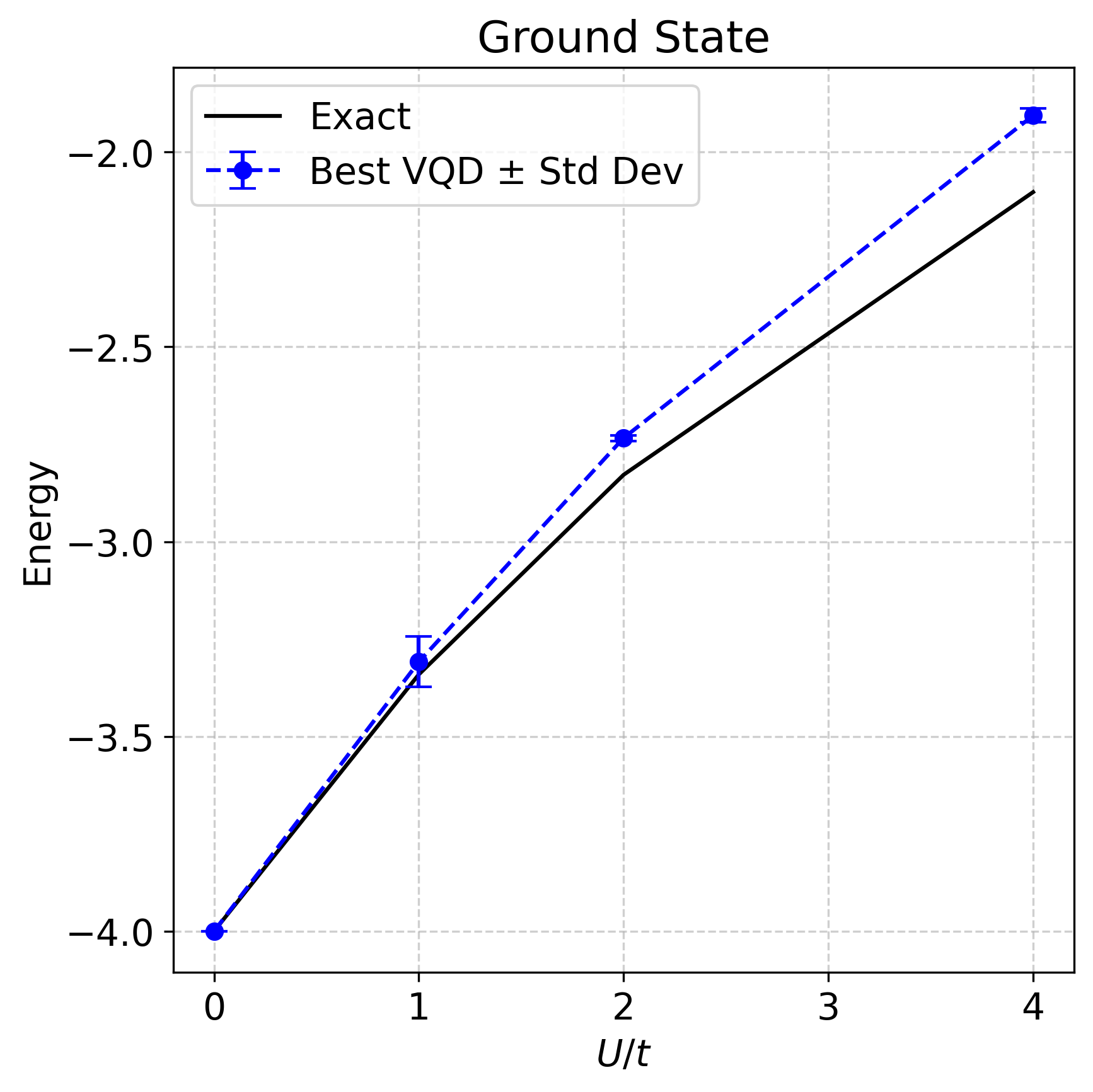}
        \caption{}
    \end{subfigure}
    \hfill
    \begin{subfigure}[t]{0.32\textwidth}
        \includegraphics[width=\linewidth]{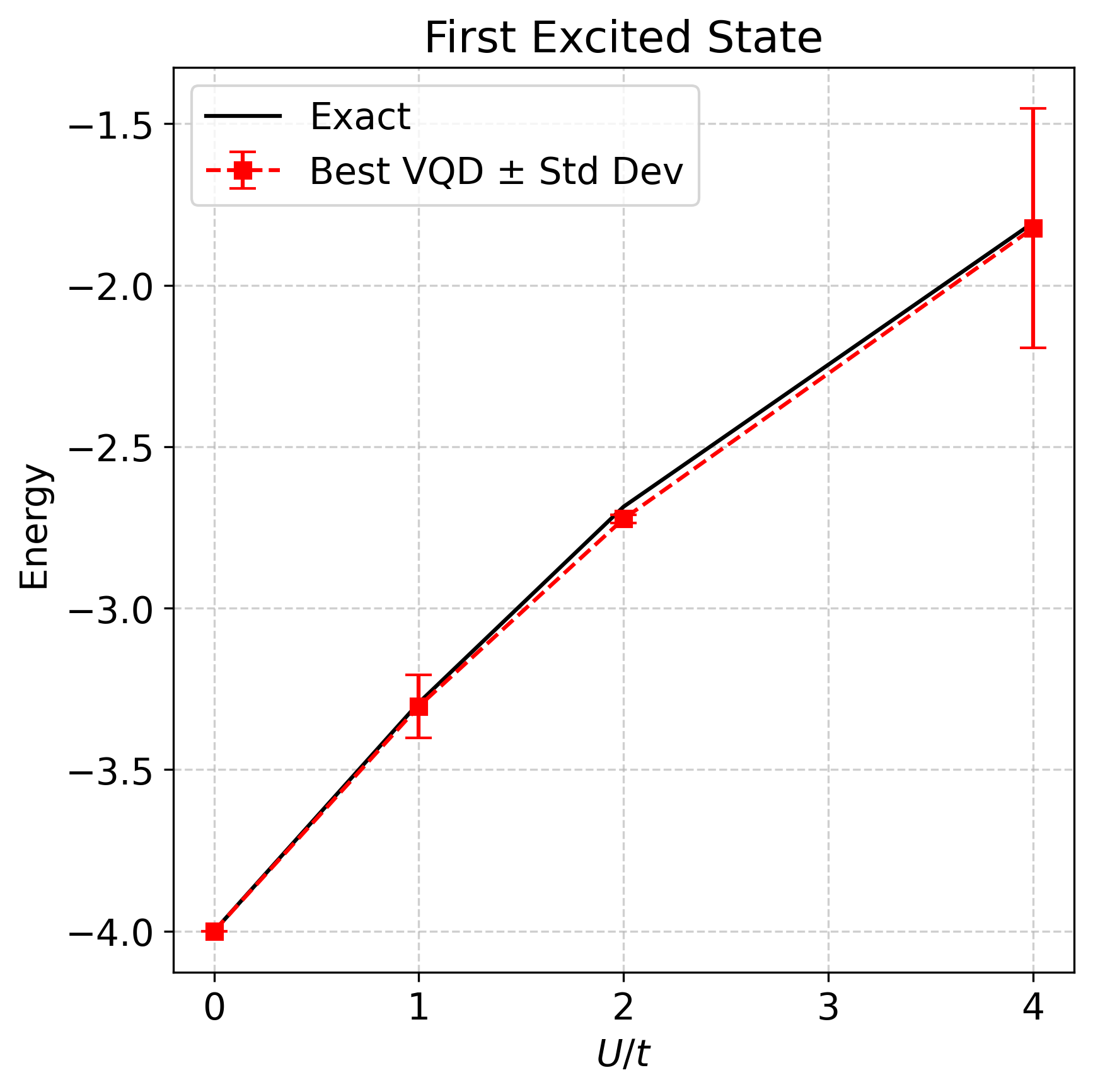}
        \caption{}
    \end{subfigure}
    \hfill
    \begin{subfigure}[t]{0.32\textwidth}
        \includegraphics[width=\linewidth]{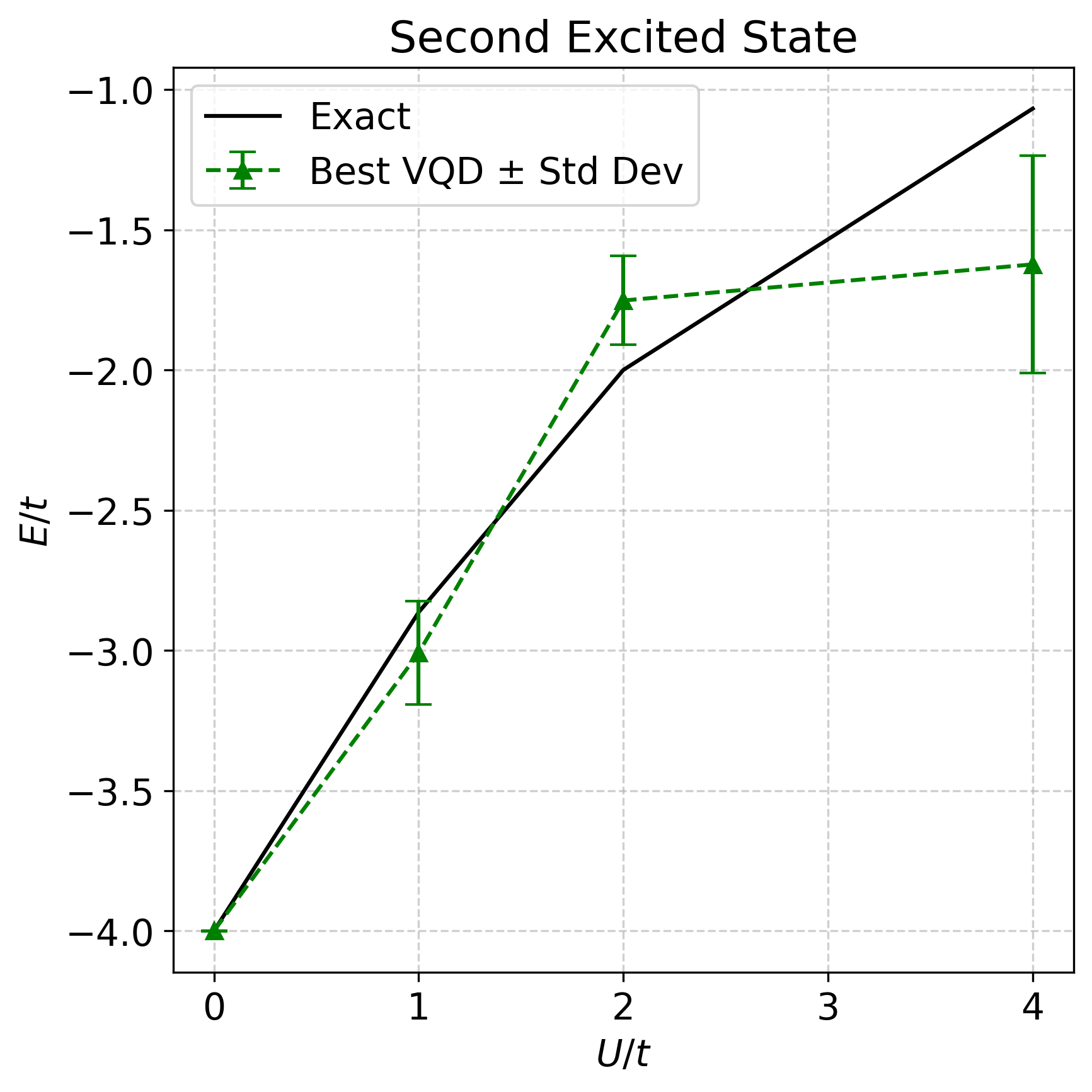}
        \caption{}
    \end{subfigure}


    \caption{(a) Ground, (b) first excited state, (c) second excited energy plotted with exact diagonalization result and VQE, with the horizontal axis U/t =0,1,2,4, for 2$\times$2 without fermionic swap.}
    \label{fig:2*2}
\end{figure*}

\begin{figure*}[t]
    \centering

    \begin{subfigure}[t]{0.32\textwidth}
        \includegraphics[width=\linewidth]{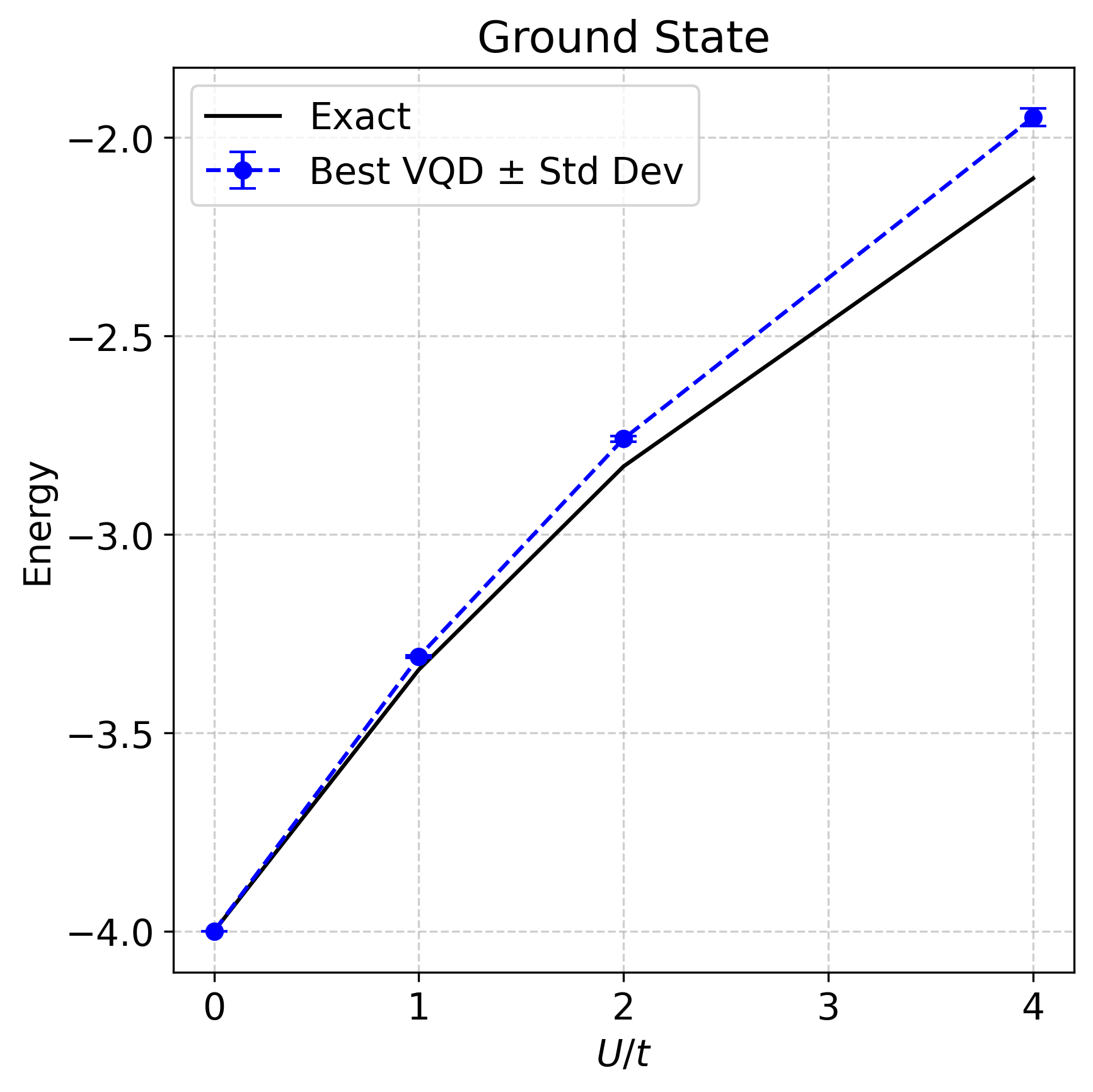}
        \caption{}
    \end{subfigure}
    \hfill
    \begin{subfigure}[t]{0.32\textwidth}
        \includegraphics[width=\linewidth]{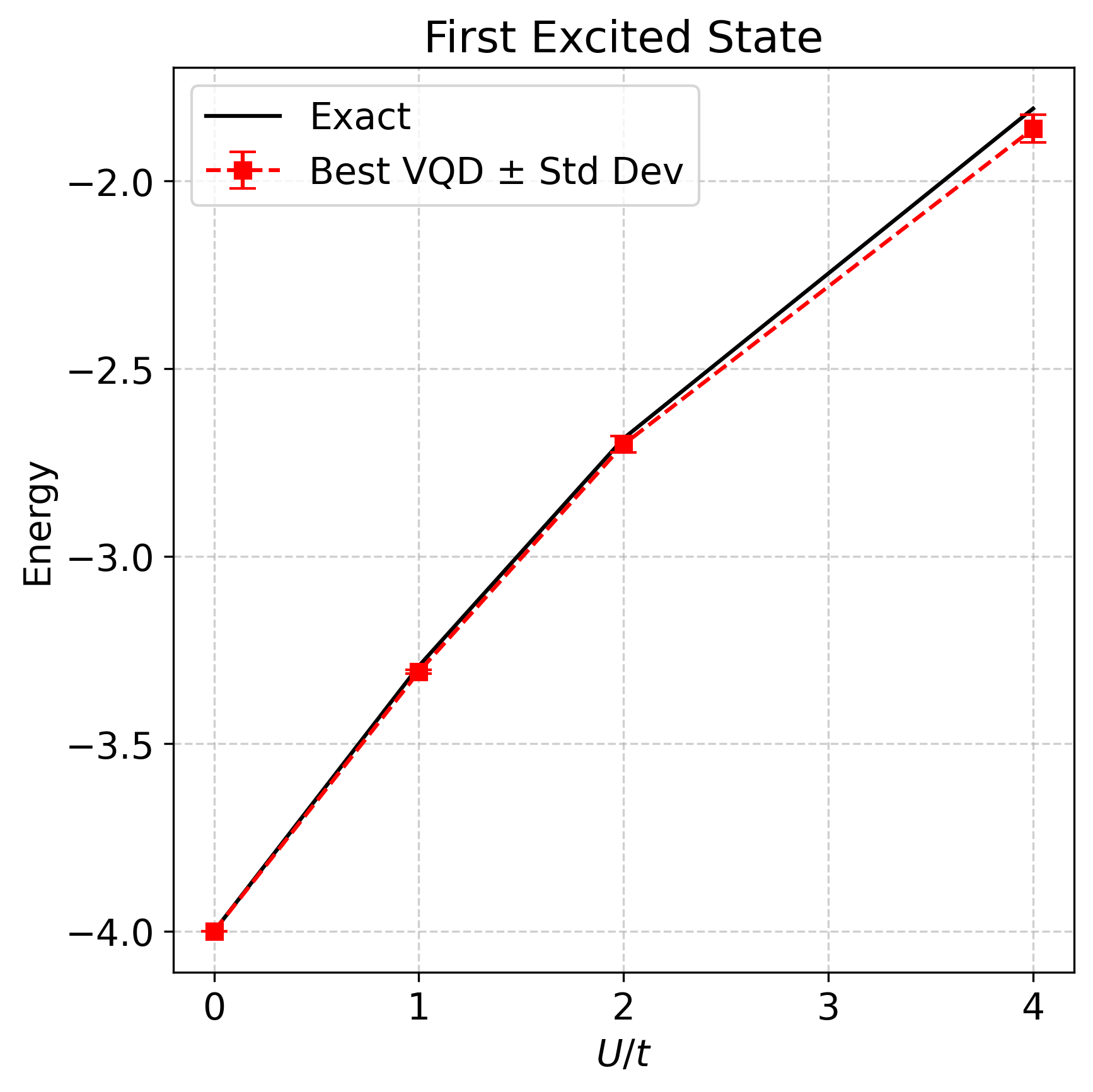}
        \caption{}
    \end{subfigure}
    \hfill
    \begin{subfigure}[t]{0.32\textwidth}
        \includegraphics[width=\linewidth]{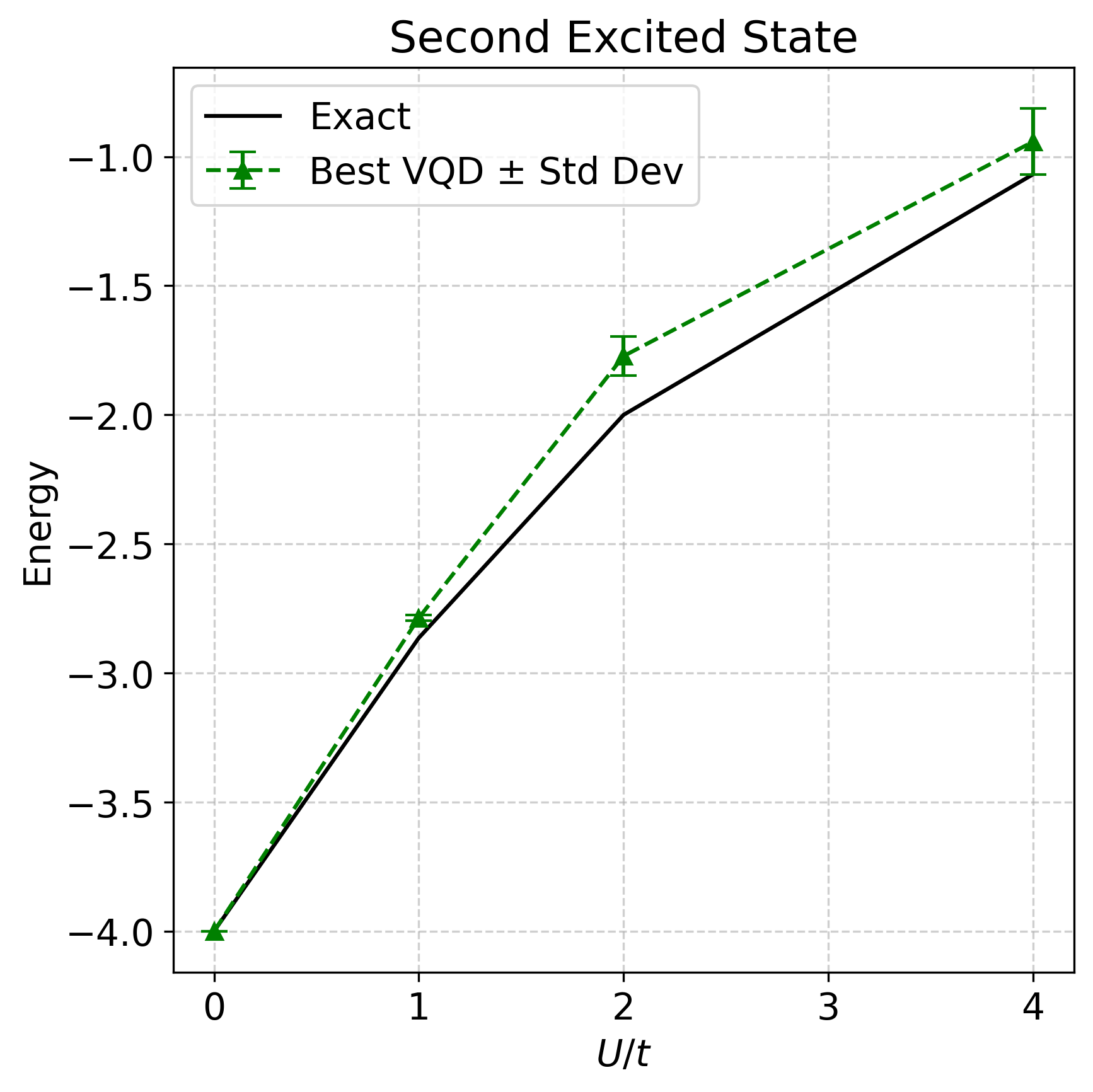}
        \caption{}
    \end{subfigure}


    \caption{(a) Ground, (b) first excited state, (c) second excited energy plotted with exact diagonalization result and VQE, with the horizontal axis U/t =0,1,2,4, for 2$\times$2 with fermionic swap}
    \label{fig:2*2f}
\end{figure*}

\begin{figure*}[t]
    \centering
    \begin{subfigure}[t]{0.32\textwidth}
        \includegraphics[width=\linewidth]{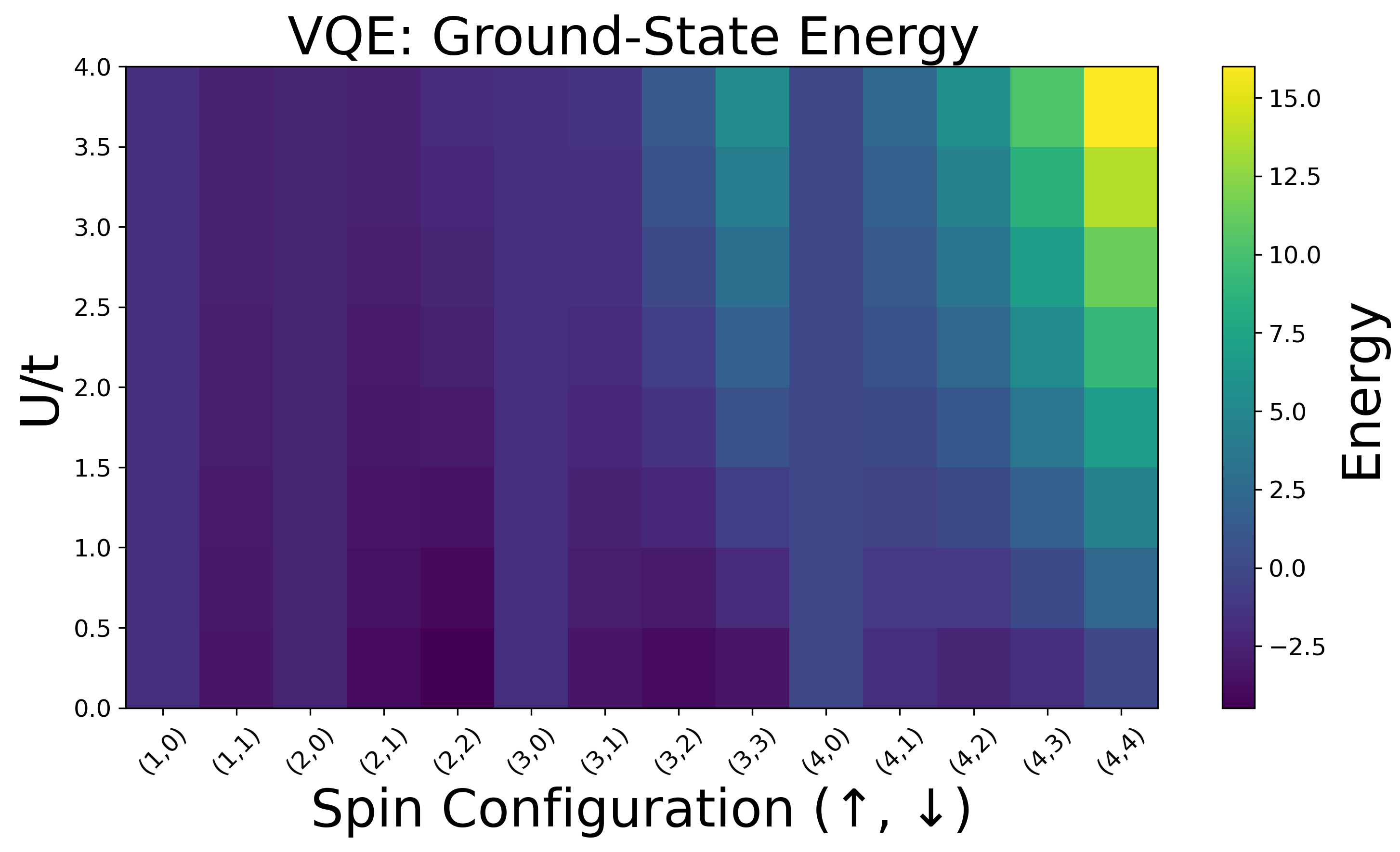}
        \caption{}
    \end{subfigure}
    \hfill
    \begin{subfigure}[t]{0.32\textwidth}
        \includegraphics[width=\linewidth]{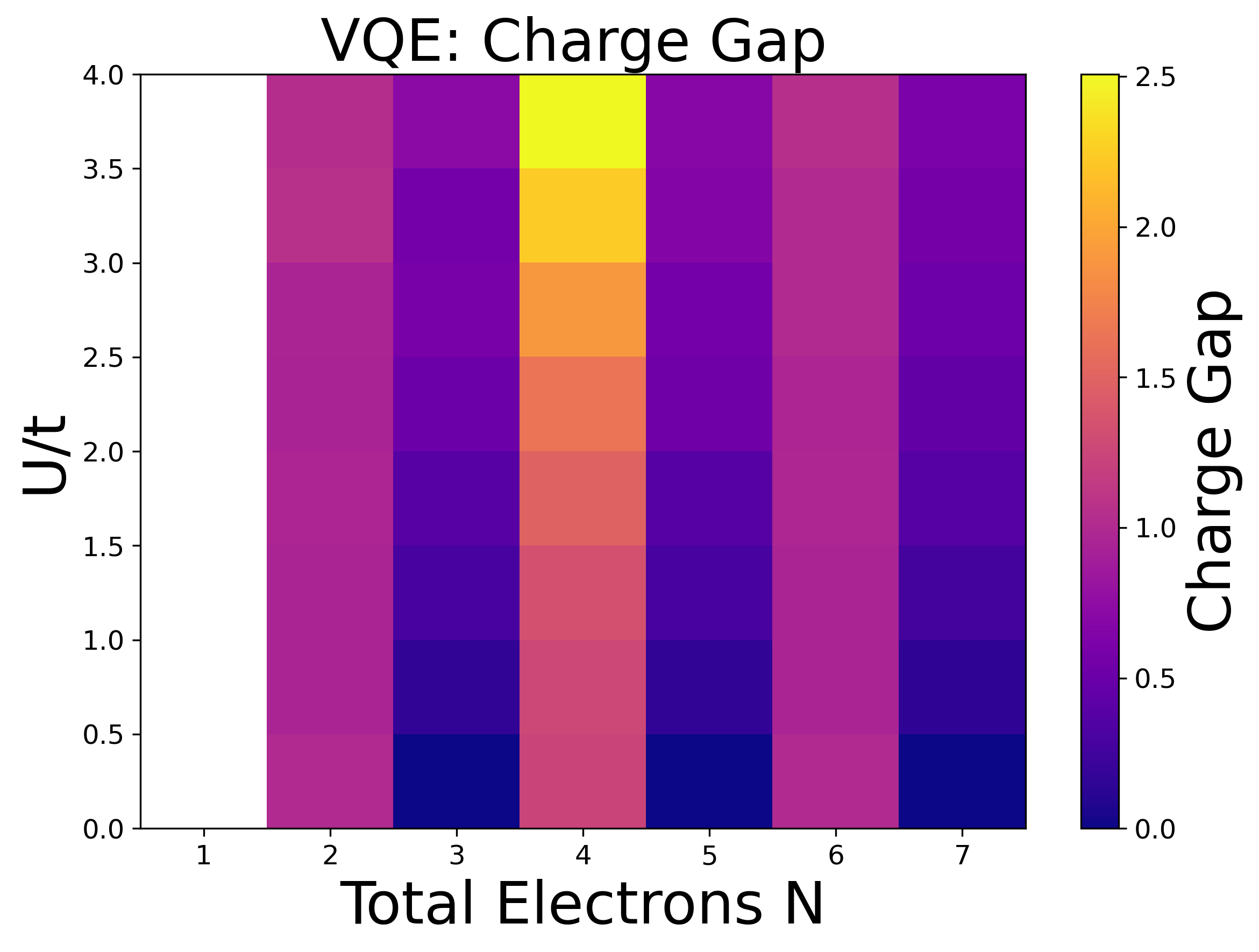}
        \caption{}
    \end{subfigure}
    \hfill
    \begin{subfigure}[t]{0.32\textwidth}
        \includegraphics[width=\linewidth]{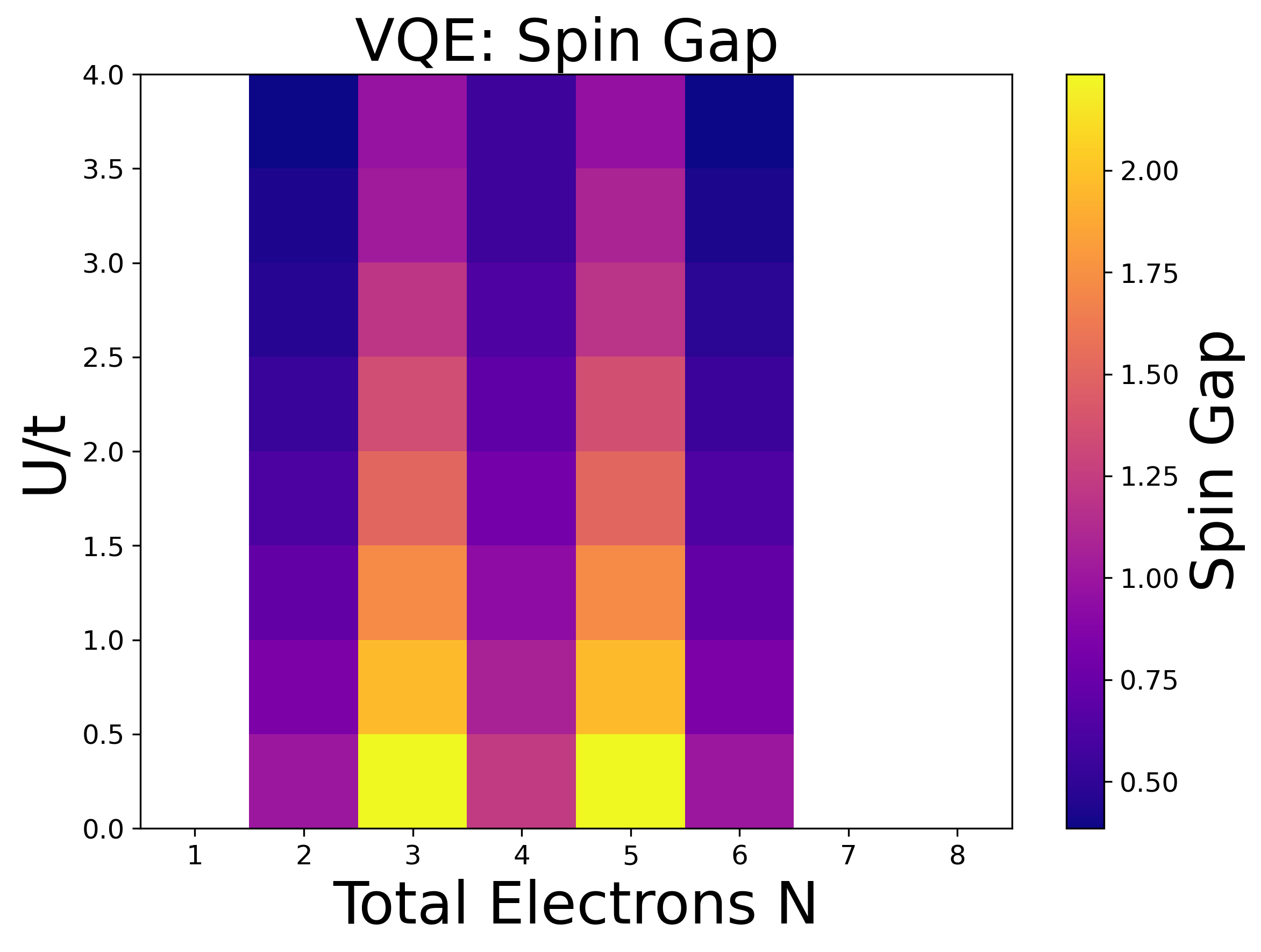}
        \caption{}
    \end{subfigure}
    \begin{subfigure}[t]{0.32\textwidth}
        \includegraphics[width=\linewidth]{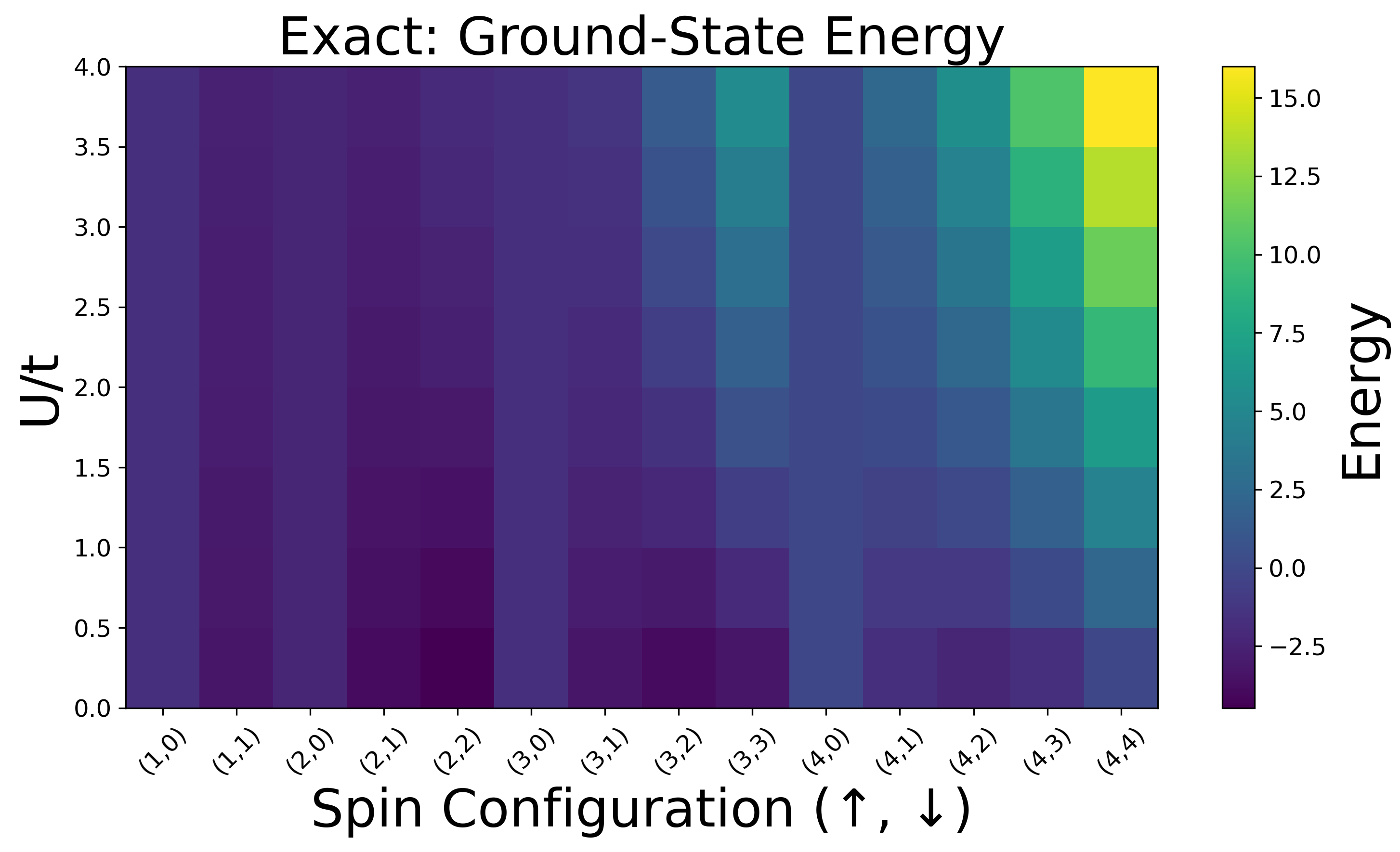}
        \caption{}
    \end{subfigure}
    \hfill
    \begin{subfigure}[t]{0.32\textwidth}
        \includegraphics[width=\linewidth]{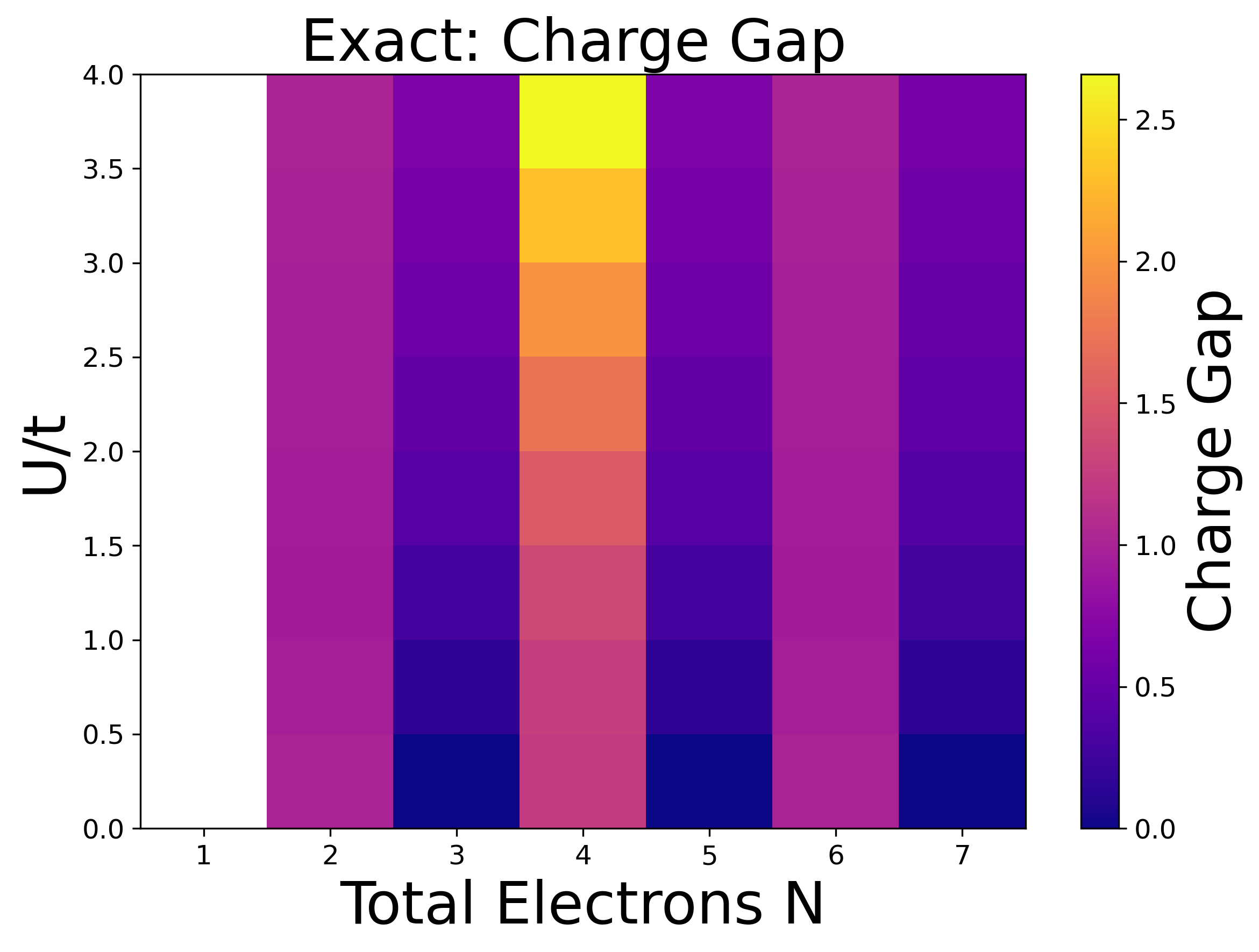}
        \caption{}
    \end{subfigure}
    \hfill
    \begin{subfigure}[t]{0.32\textwidth}
        \includegraphics[width=\linewidth]{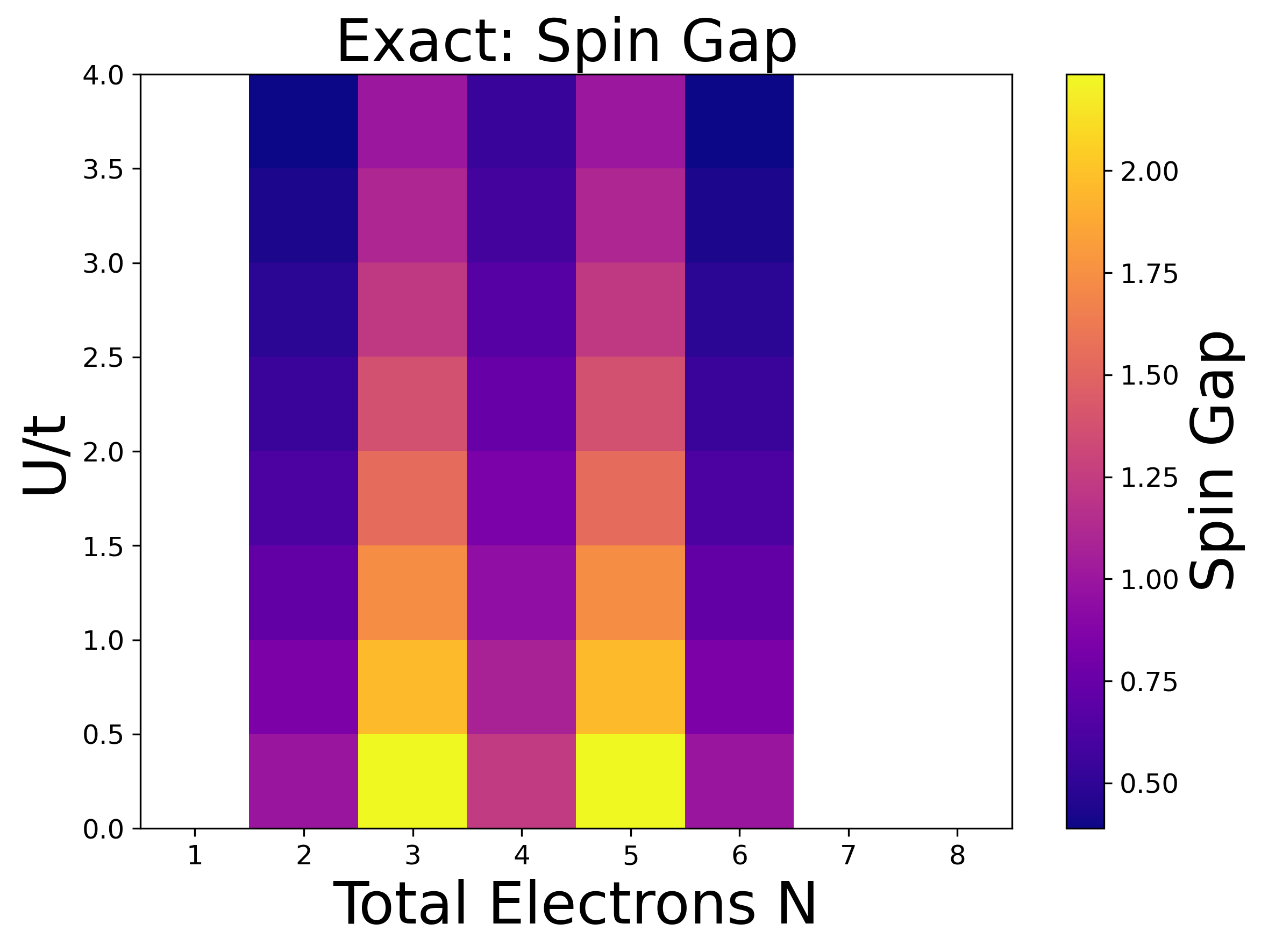}
        \caption{}
    \end{subfigure}
    
    \caption{Different energy gaps showcased for 4$\times$1 Hubbard model with electron configuration in the x-axis and U values in the y-axis for ground state energy (a) VQE and (d) exact, and energy gaps for (b) VQE charge, (c) VQE spin, (e) exact charge, and (f) exact spin, electron numbers are in the x-axis and U in the y-axis.}
    \label{fig:phasediagrams8}
\end{figure*}

\begin{figure*}[t]
    \centering
    
    \begin{subfigure}[t]{0.32\textwidth}
        \includegraphics[width=\linewidth]{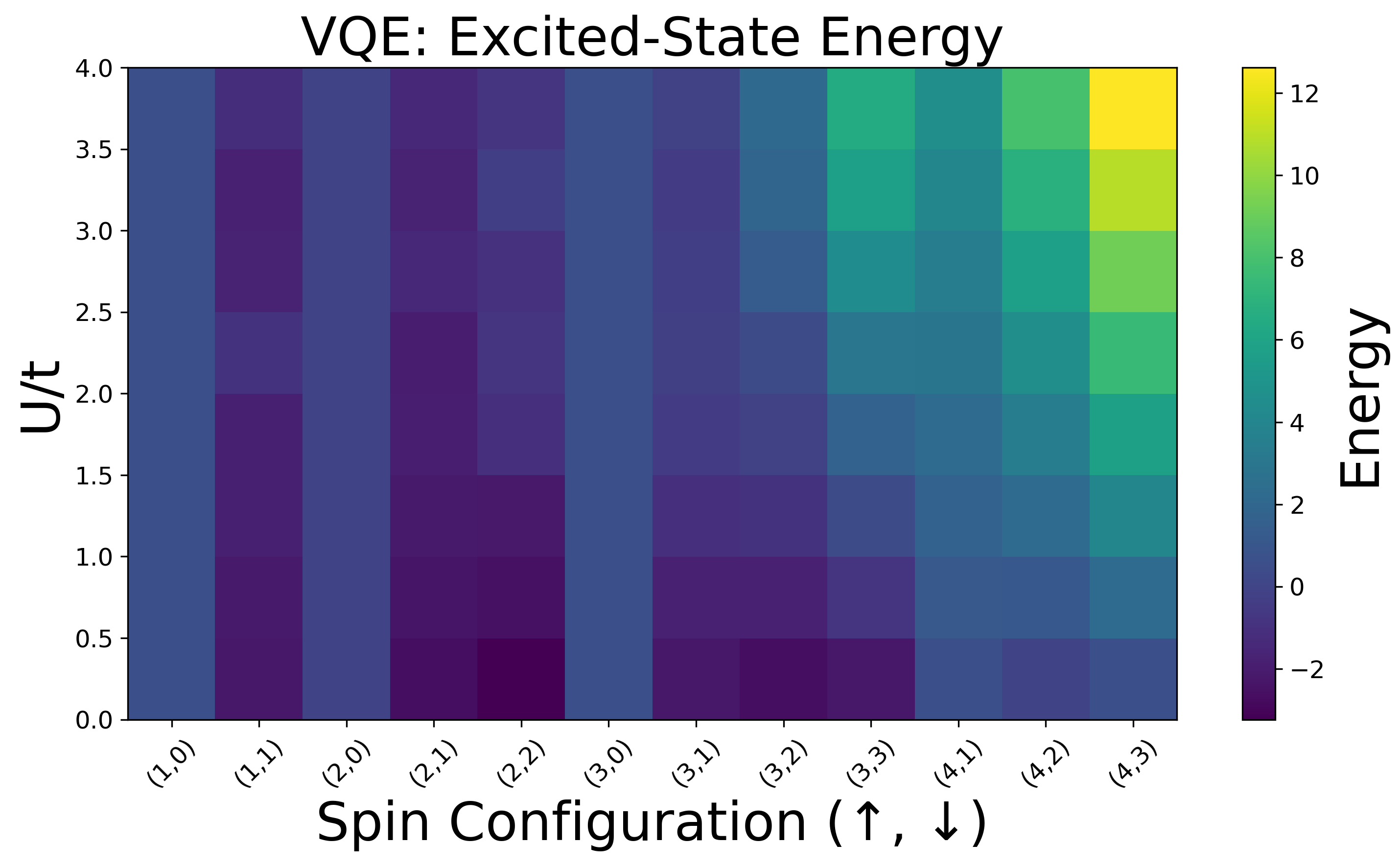}
        \caption{}
    \end{subfigure}
    \hfill
    \begin{subfigure}[t]{0.32\textwidth}
        \includegraphics[width=\linewidth]{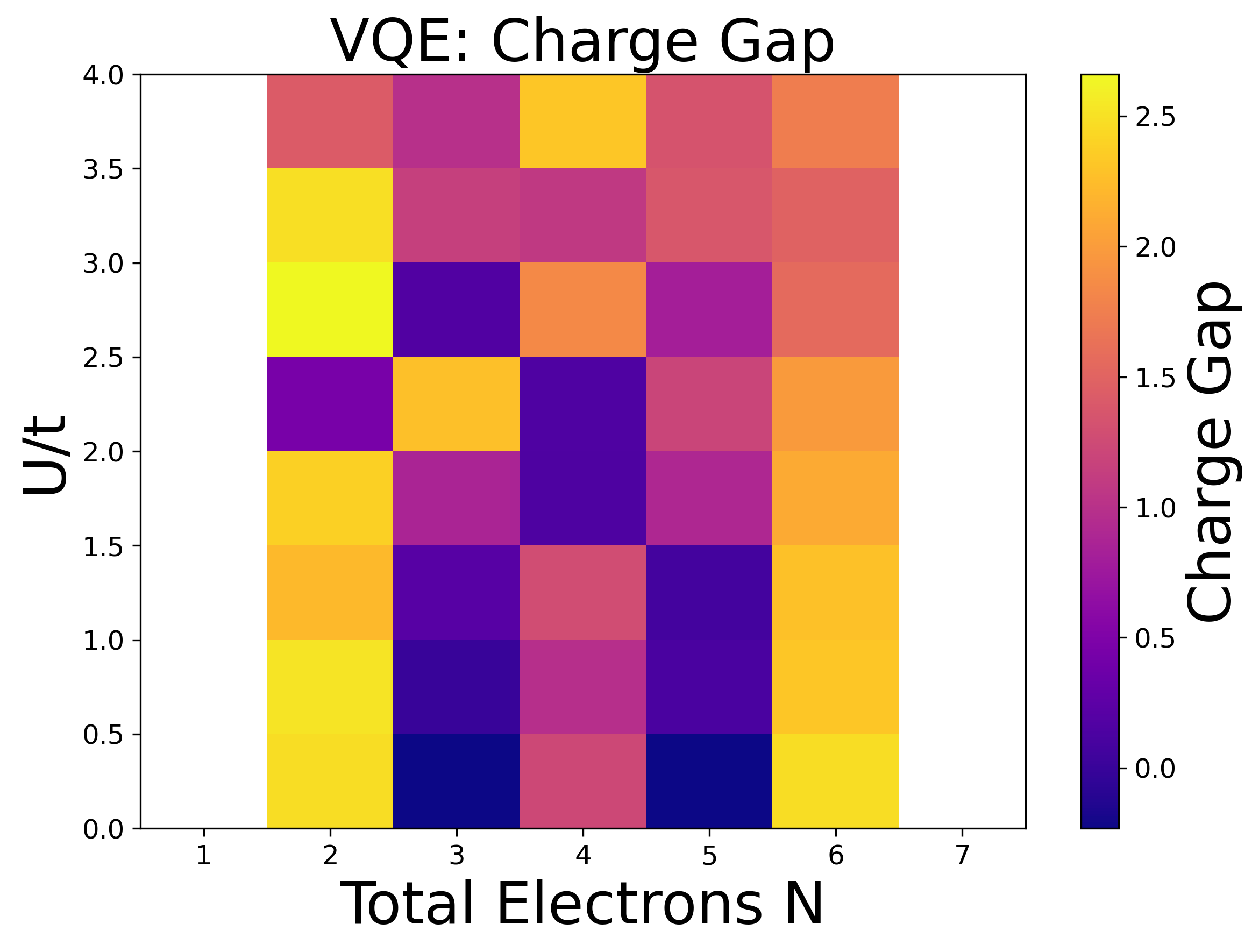}
        \caption{}
    \end{subfigure}
    \hfill
    \begin{subfigure}[t]{0.32\textwidth}
        \includegraphics[width=\linewidth]{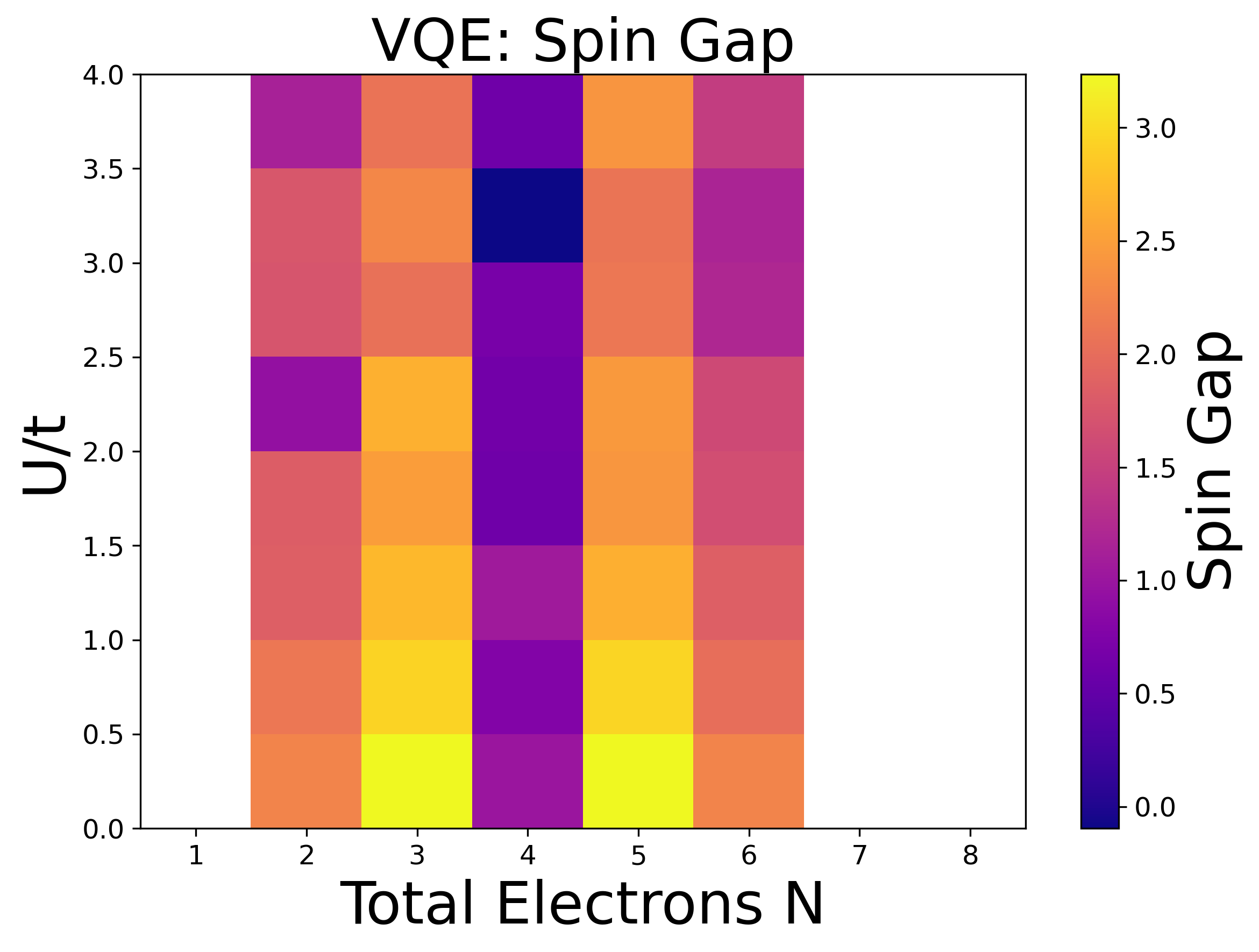}
        \caption{}
    \end{subfigure}
    
    \begin{subfigure}[t]{0.32\textwidth}
        \includegraphics[width=\linewidth]{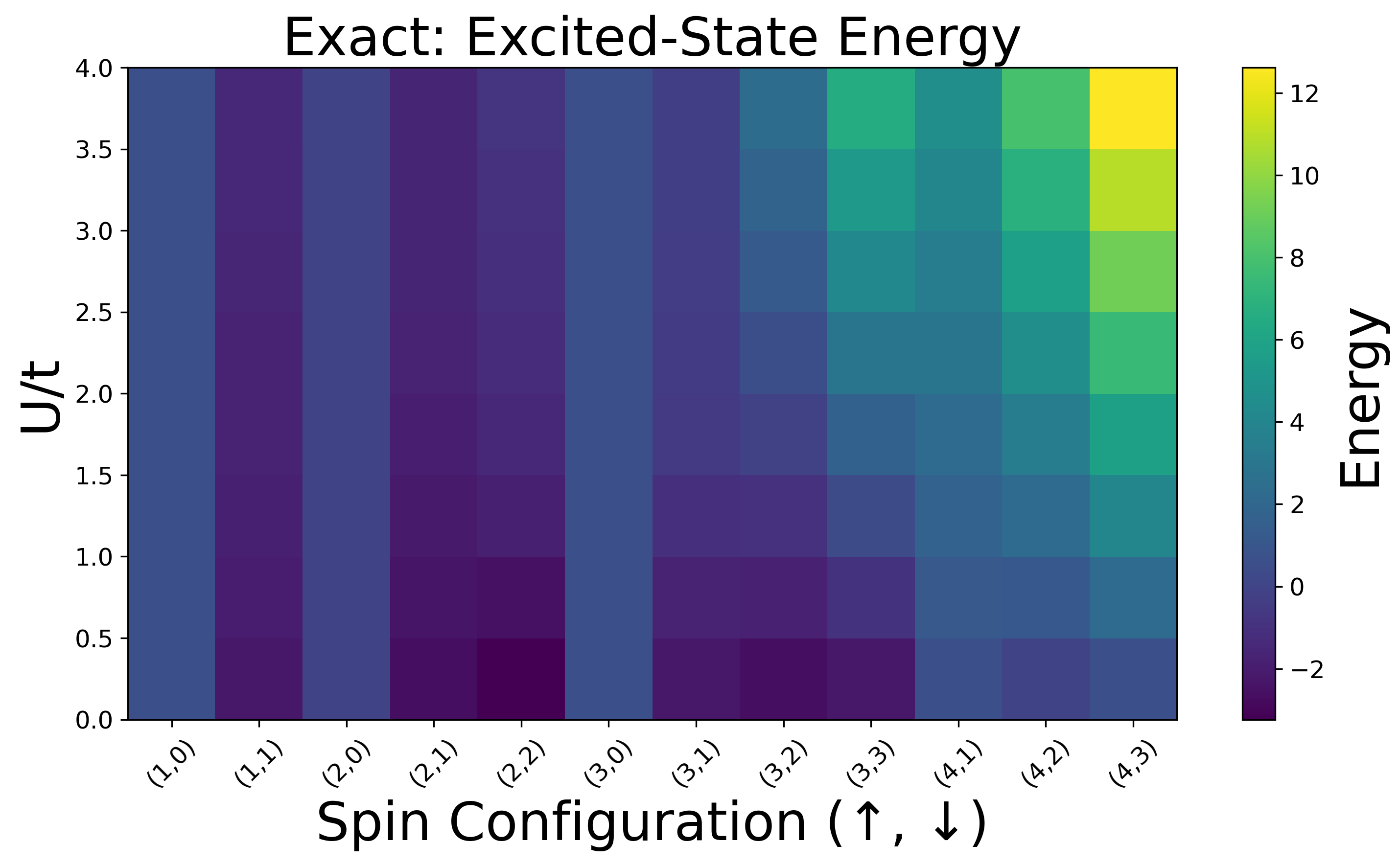}
        \caption{}
    \end{subfigure}
    \hfill
    \begin{subfigure}[t]{0.32\textwidth}
        \includegraphics[width=\linewidth]{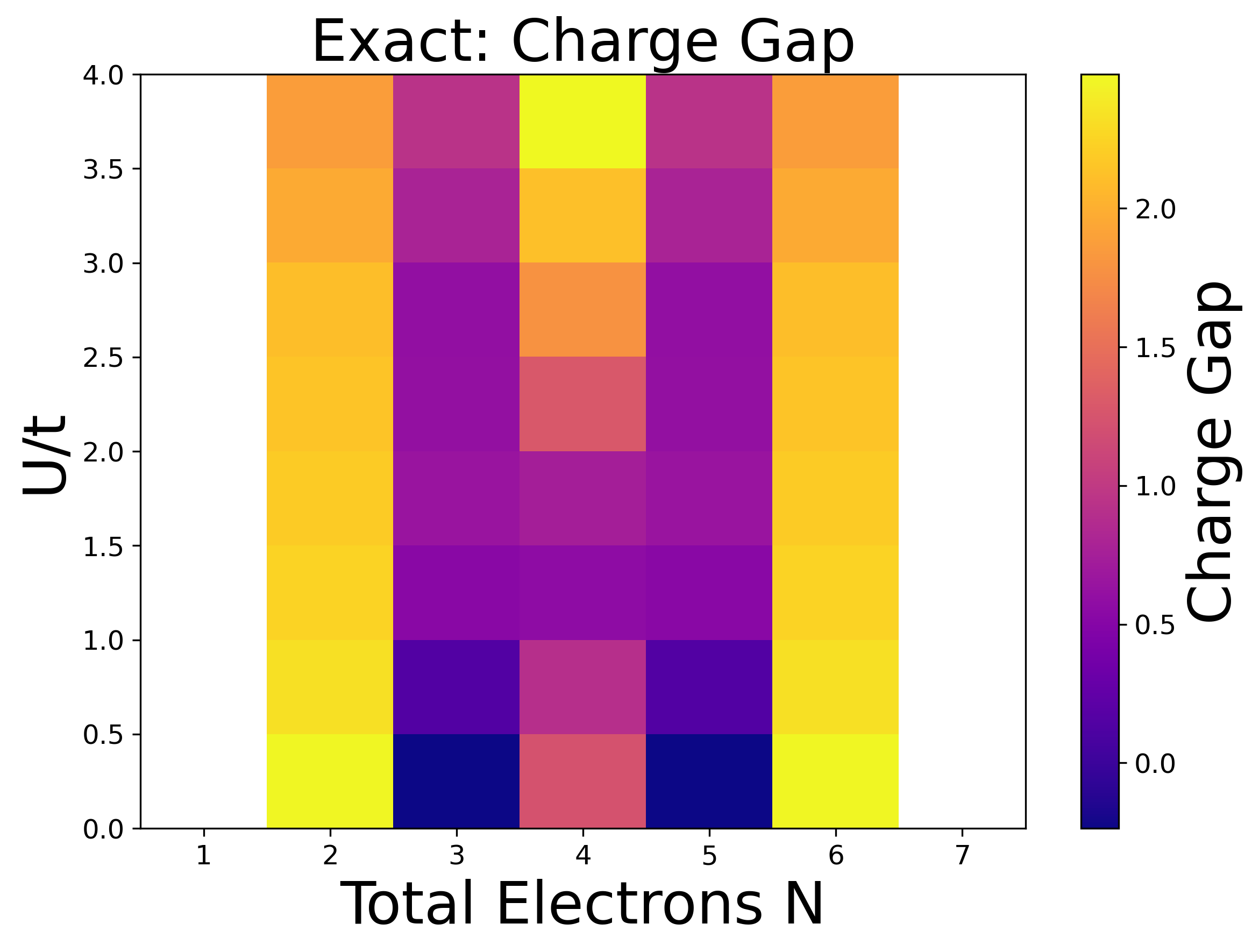}
        \caption{}
    \end{subfigure}
    \hfill
    \begin{subfigure}[t]{0.32\textwidth}
        \includegraphics[width=\linewidth]{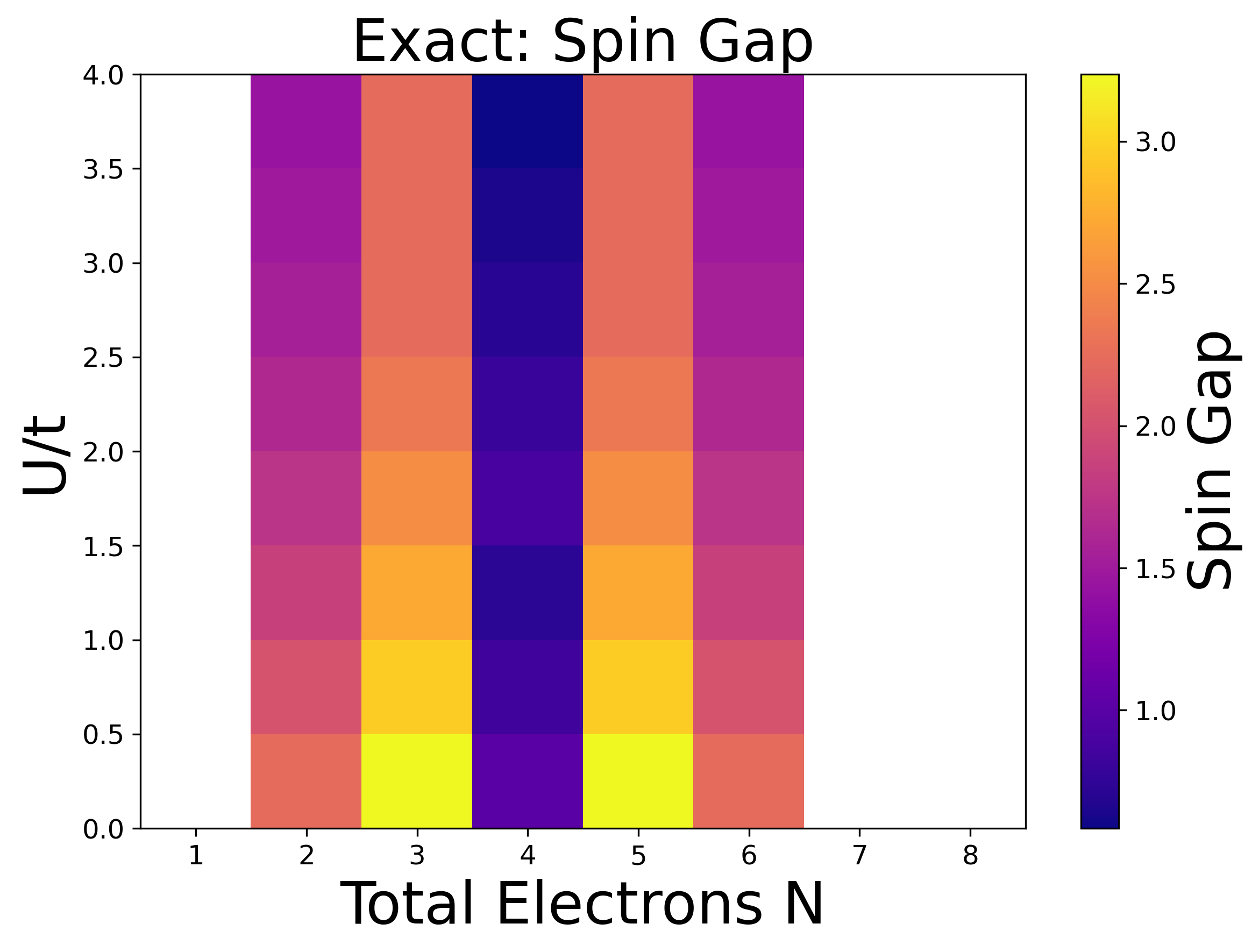}
        \caption{}
    \end{subfigure}
    \caption{Different energy gaps showcased for the 4$\times$1 Hubbard model with electron configuration in the x-axis and U values in the y-axis for the 2nd excited energy (a) VQE and (d) exact, and for (b) VQE charge gap, (c) VQE spin gap, (e) exact charge gap, and (f) exact spin gap, the electron numbers are in the x-axis and U in the y-axis.}
    \label{fig:phasediagrams9}
\end{figure*}

\begin{figure*}
    \centering
    \begin{subfigure}[t]{0.32\textwidth}
        \includegraphics[width=\linewidth]{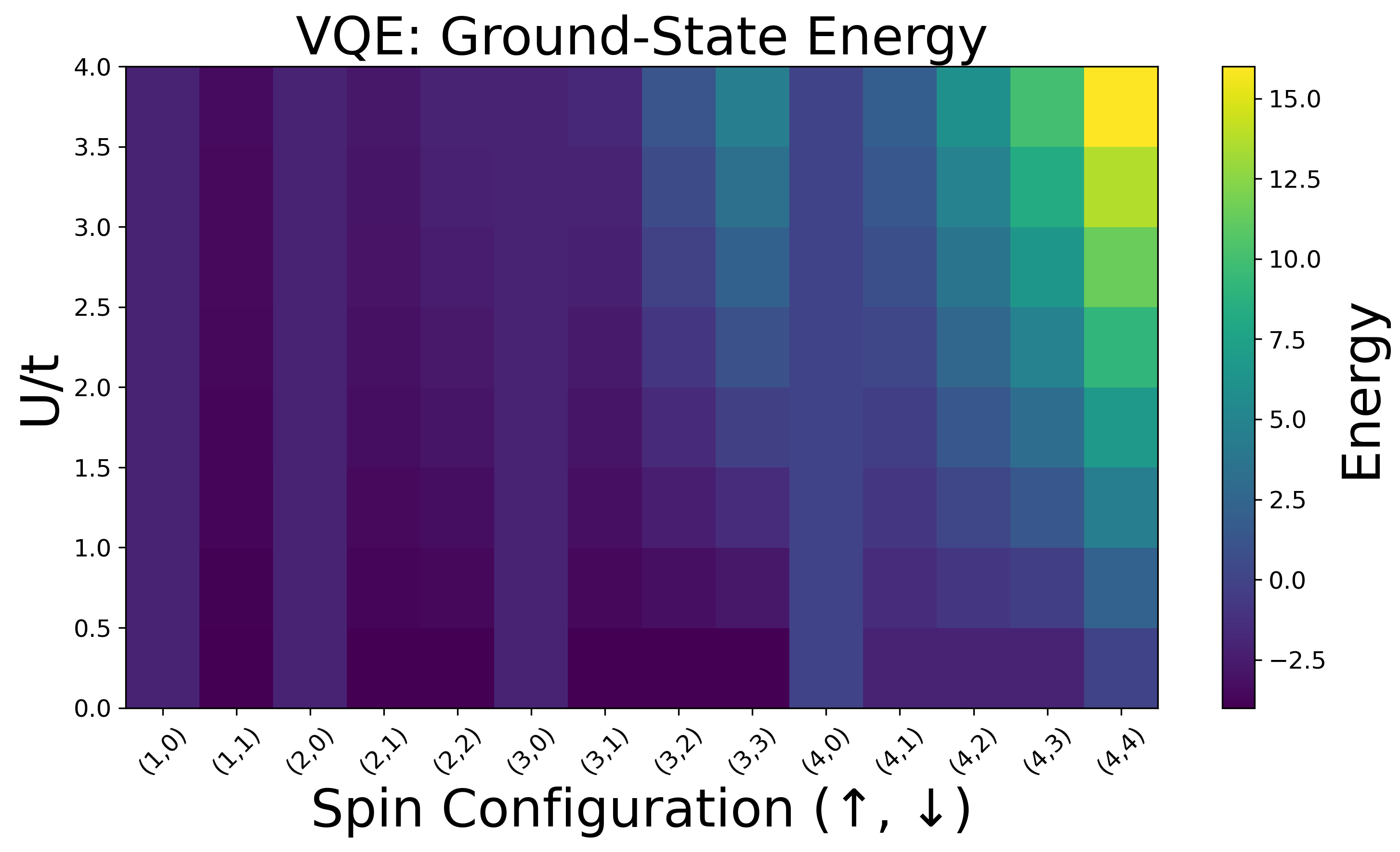}
        \caption{}
    \end{subfigure}
    \hfill
    \begin{subfigure}[t]{0.32\textwidth}
        \includegraphics[width=\linewidth]{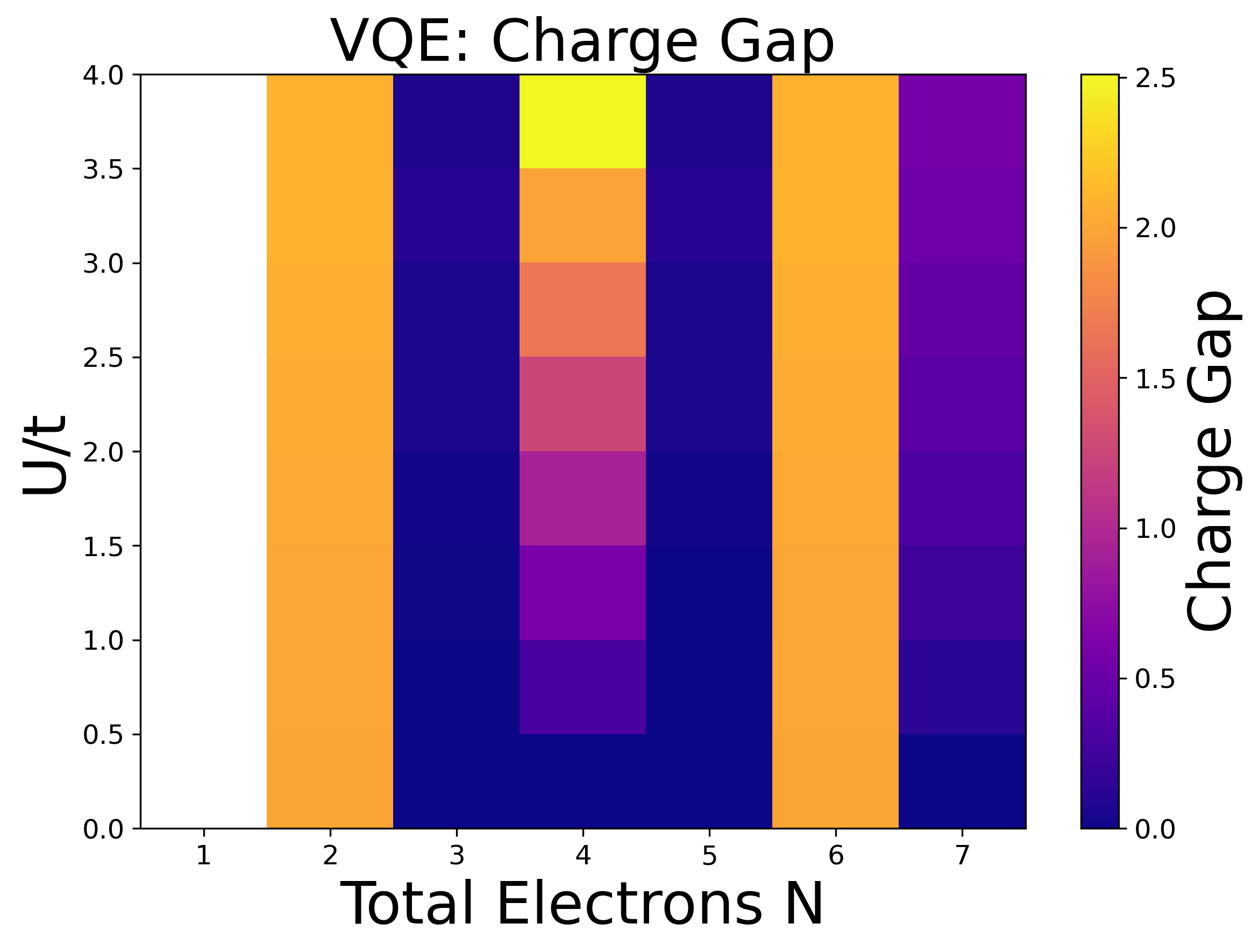}
        \caption{}
    \end{subfigure}
    \hfill
    \begin{subfigure}[t]{0.32\textwidth}
        \includegraphics[width=\linewidth]{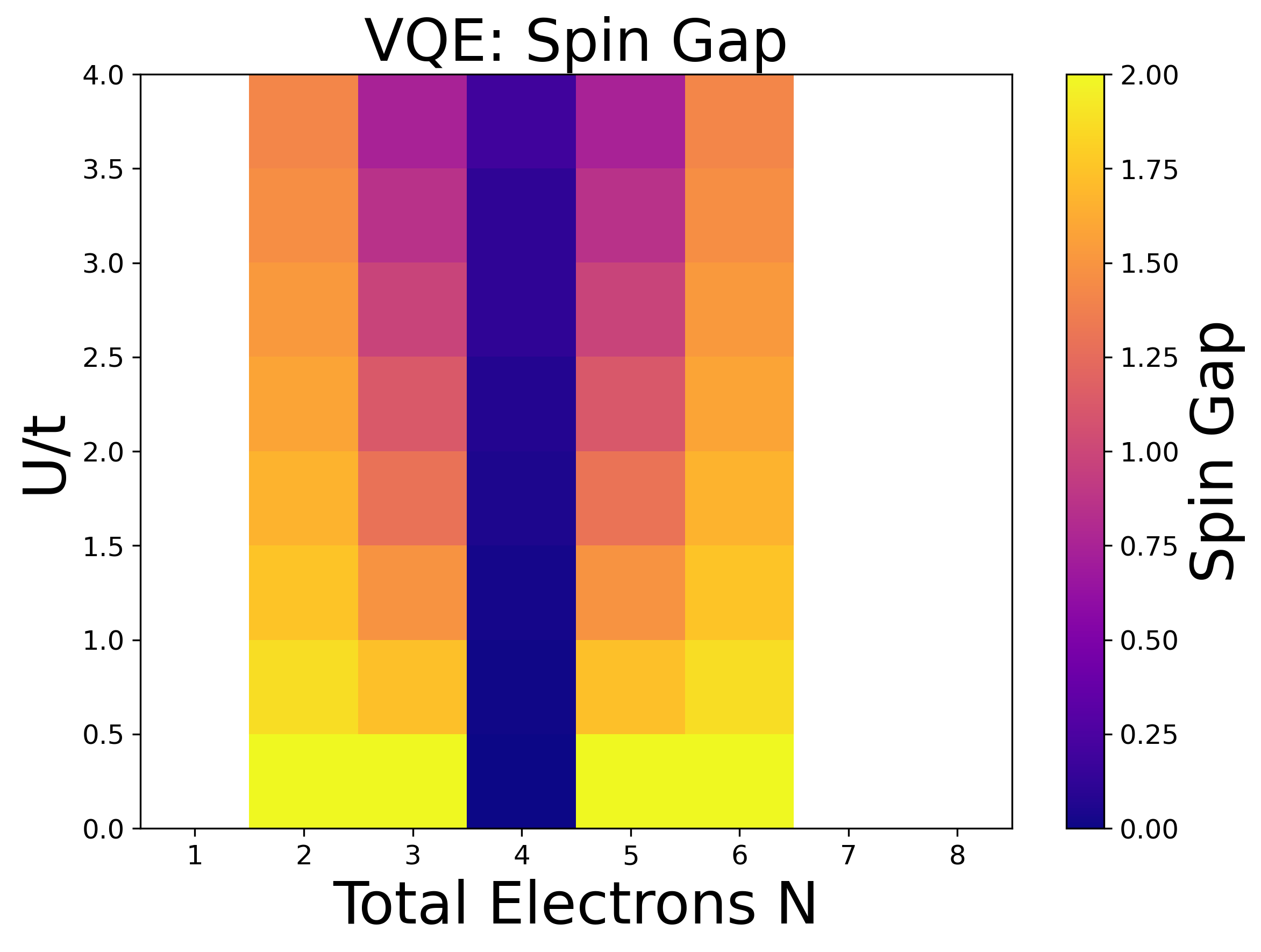}
        \caption{}
    \end{subfigure}
    \begin{subfigure}[t]{0.32\textwidth}
        \includegraphics[width=\linewidth]{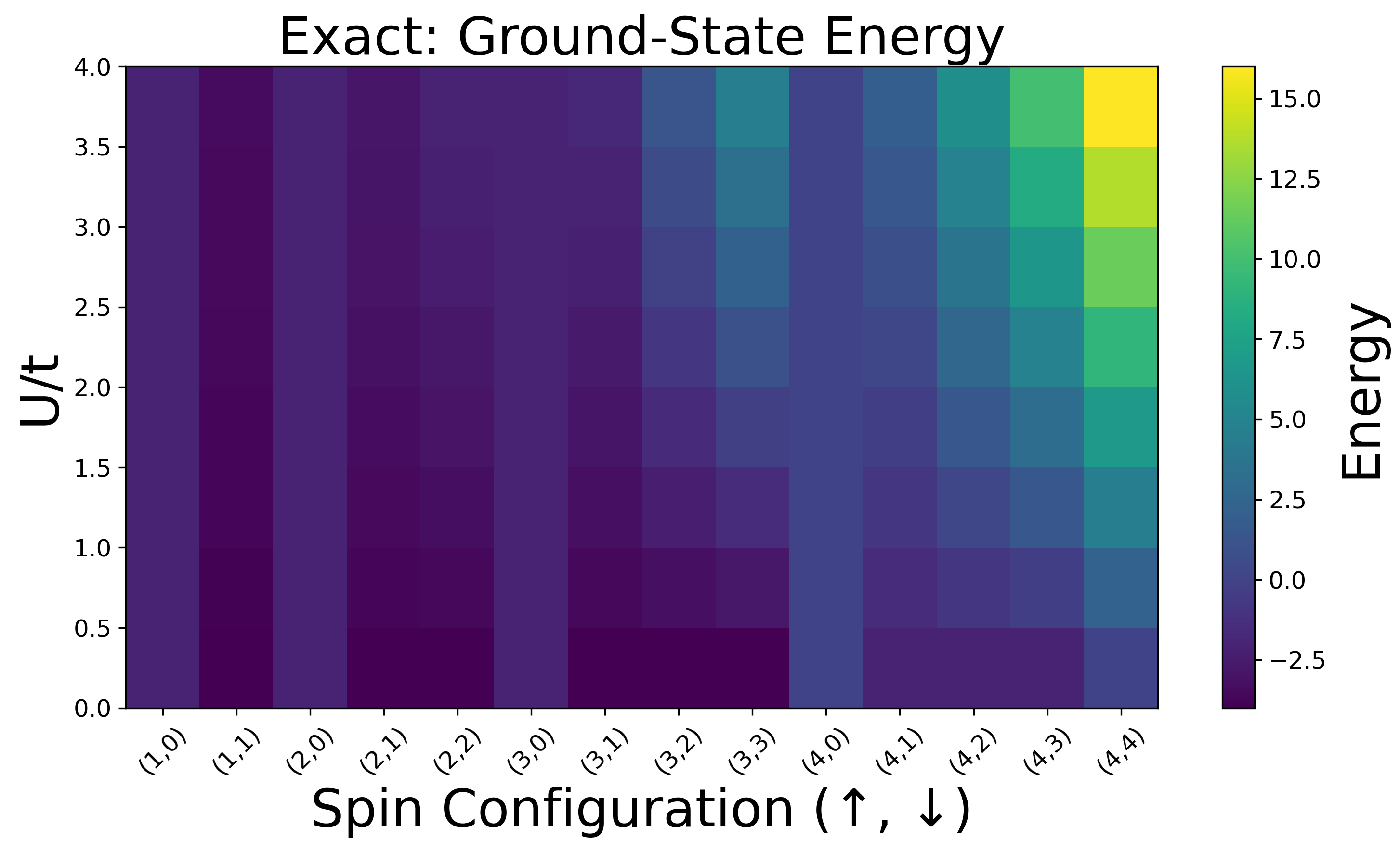}
        \caption{}
    \end{subfigure}
    \hfill
    \begin{subfigure}[t]{0.32\textwidth}
        \includegraphics[width=\linewidth]{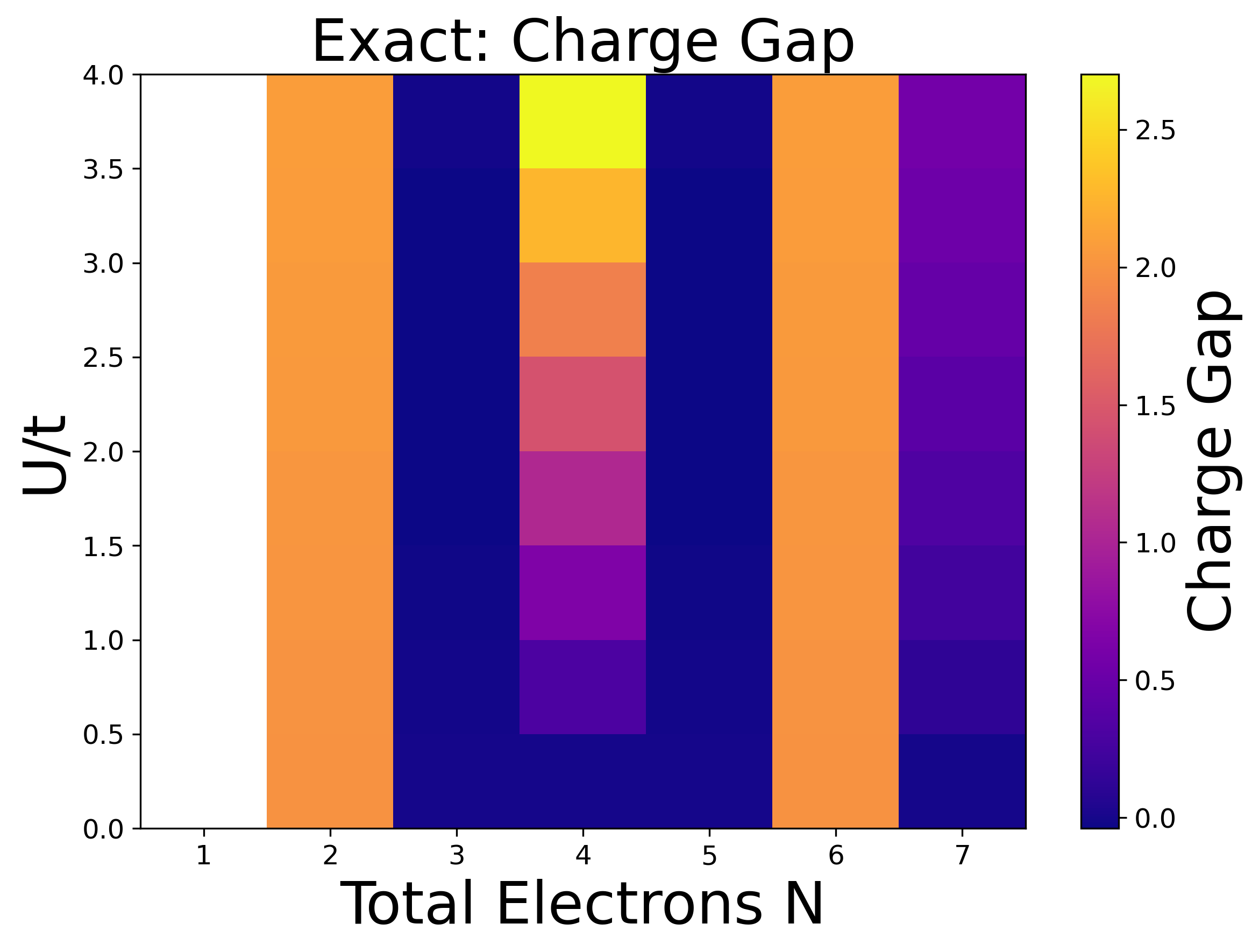}
        \caption{}
    \end{subfigure}
    \hfill
    \begin{subfigure}[t]{0.32\textwidth}
        \includegraphics[width=\linewidth]{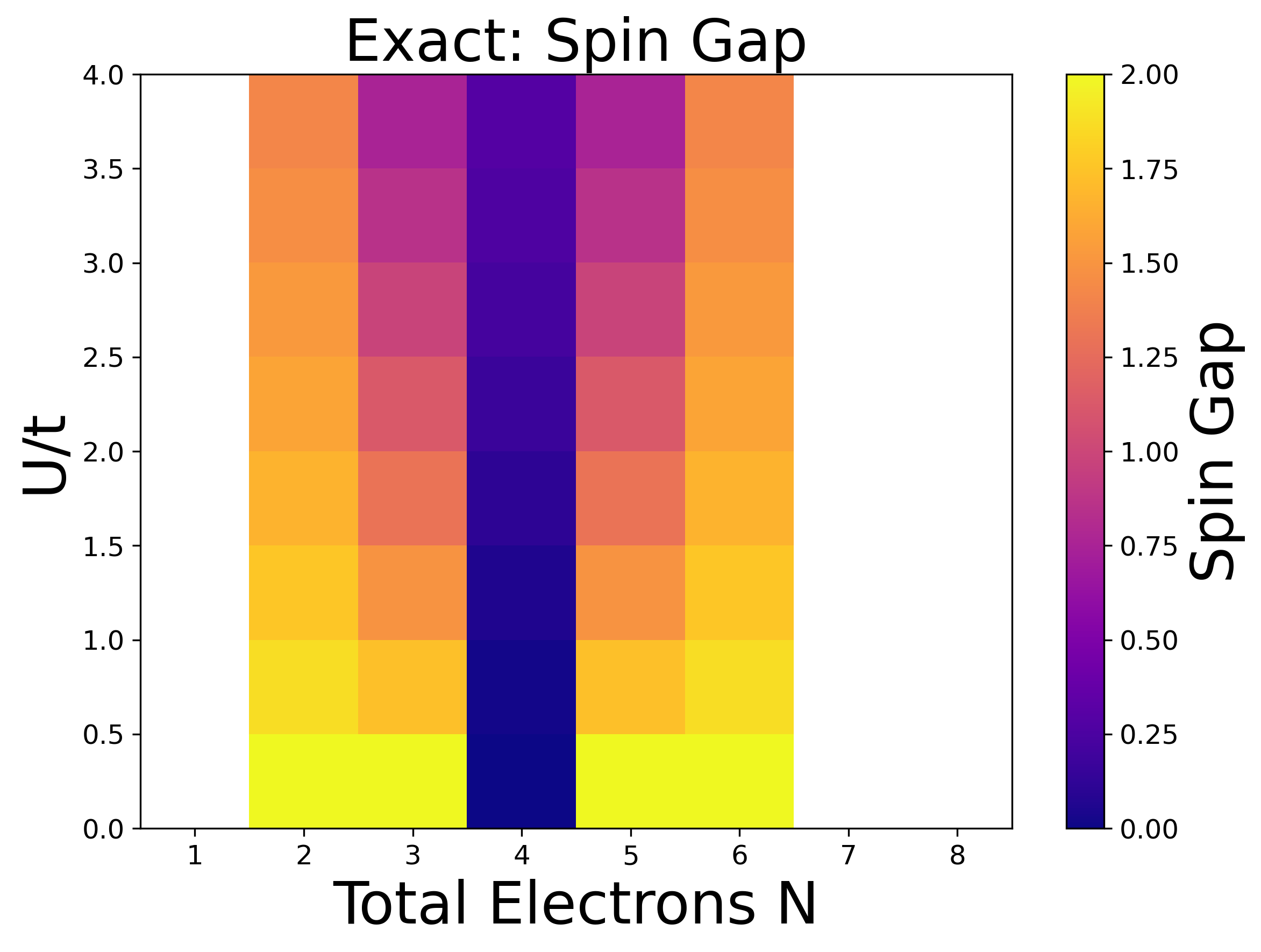}
        \caption{}
    \end{subfigure}
    \caption{Different energy gaps showcased for the 2$\times$2 Hubbard model with electron configuration in the x-axis and U values in the y-axis for the ground energy (a) VQE and (d) exact, and for (b) VQE charge gap, (c) VQE spin gap, (e) exact charge gap, and (f) exact spin gap, the electron numbers are in the x-axis and U in the y-axis.}
    \label{f :phasediagrams10}
\end{figure*}

\subsection{Comparison of VQE results with exact energies}
The ground, first, and second excited state energies of different configurations were computed using the VQE and compared with the exact energies obtained through full diagonalization. It is observed that the energies increase as the Coulombic repulsion \(U/t\) increases for all configurations. There is also an increase in error from the ground state to the excited states, as excited states are more difficult to obtain accurately than  the ground state. It is further observed that the change in energy from the ground state to the excited states is more pronounced for the \(4 \times 1\) Fig. \ref{fig:4*1} configuration than for the \(2 \times 2\) configuration Fig. \ref{fig:2*2f}. As seen in Fig.~\ref{fig:2*2f}, the energy of the \(2 \times 2\) system barely changes until the potential exceeds \(U/t = 2\). Fig.~\ref{fig:4*1} shows the exact and VQE energies for linear \(4 \times 1\) chains, which involve only horizontal hopping and therefore do not require fermionic swaps in their ansatz. Two-dimensional models with a \(2 \times 2\) lattice have also been studied (Fig.~\ref{fig:2*2f}), where fermionic swaps are required to account for non-adjacent vertical hopping. It is observed that without fermionic swaps (Fig.~\ref{fig:2*2}), the error is higher, as the non-adjacent hopping is not properly represented in the ansatz circuit. This highlights the importance of using an ansatz that preserves the structure of the Hamiltonian. The results demonstrate the sensitivity of the ansatz: one cannot use an arbitrary ansatz, and incorporating the Hamiltonian's structure into the ansatz plays a crucial role. Although the HVA currently lacks formal proof, this work demonstrates the importance of leveraging the Hamiltonian structure in the ansatz circuit, thereby strengthening the case for HVA.
Now, proceeding with the detailed analysis of each figure subsequently.
Fig.~\ref{fig:4*1} compares the exact diagonalization results with the energies obtained using the VQE method for the $4 \times 1$ Fermi-Hubbard lattice at interaction strengths $U/t = 0, 1, 2, 4$. Subfigure (a) shows the ground state energy. At $U/t = 0$, both exact and VQE results give $E_0 \approx -4.50$, with negligible deviation. For $U/t = 1$, the energies are $-3.56$ (exact) and $-3.55$ (VQE), maintaining a difference below $0.01$. Even at $U/t = 4$, the VQE energy of $-1.98$ closely matches the exact $-2.02$, demonstrating sub-2\% relative error for the ground state across all interaction strengths. Subfigure (b) presents the first excited state energy. At $U/t = 0$, the VQE and exact results are nearly identical at $-3.20$. A small overestimation appears for $U/t = 1$ with VQE at $-2.40$ compared to the exact $-2.55$, giving an error of $\sim 0.15$. At $U/t = 2$, VQE yields $-1.85$ versus the exact $-2.10$, and at $U/t = 4$, the energies are $-1.15$ (VQE) and $-1.45$ (exact), showing a maximum deviation of approximately $0.3$. These differences increase with larger $U/t$, reflecting the growing difficulty of accurately capturing excited states with shallow circuits. Subfigure (c) shows the second excited state energy. At $U/t = 0$, the VQE result of $-3.18$ aligns well with the exact $-3.15$. However, the deviation becomes significant for stronger interactions. At $U/t = 1$, the VQE energy is $-2.25$, overshooting the exact $-2.10$ by $0.15$. At $U/t = 2$, the VQE result of $-1.05$ differs slightly from the exact $-1.25$, while for $U/t = 4$, the second excited state energy exact value is -1.0 and VQE estimates it to be -0.9 with a 0.1 difference. The error bars increase for higher excited states, indicating greater optimization sensitivity and circuit-depth limitations in NISQ-era hardware. Overall, the results show that the proposed ansatz and hybrid COBYLA-LBFGS optimization achieve near-exact accuracy for the ground state, maintain moderate accuracy for the first excited state (maximum deviation $\sim 0.3$), and provide qualitatively correct but less precise estimates for the second excited state at higher $U/t$ values.

Fig.~\ref{fig:2*2} compares the exact diagonalization (Exact) and VQE Best results for the ground, first excited, and second excited state energies of a $2\times 2$ Hubbard model without fermionic swap for $U/t=0,1,2,4$. For the ground state, VQE closely matches the exact results across all $U/t$ values, starting from $-4.0$ at $U/t=0$ and increasing to approximately $-1.9$ at $U/t=4$, slightly overestimating compared to the exact value of about $-2.1$. The first excited state energies also show excellent agreement, increasing monotonically from $-4.0$ to around $-1.8$, with only minor deviations and slightly larger error bars at $U/t=4$. For the second excited state, deviations become more pronounced as $U/t$ increases: while both methods give $-4.0$ at $U/t=0$, at $U/t=4$ the exact energy reaches about $-1.0$, whereas VQE underestimates it to roughly $-1.4$ with larger uncertainties. Overall, VQE effectively reproduces exact results for the ground and first excited states, while accuracy decreases and uncertainties grow for the second excited state in the strongly interacting regime.

Fig.~\ref{fig:2*2f} shows the comparison between exact diagonalization results and VQE with fermionic swap estimates for the \(2\times2 \) Fermi-Hubbard model at \(U/t = 0, 1, 2, 4\). Subfig.~(a) depicts the ground state energy, which closely follows the exact curve with deviations below \(0.01\) for \(U/t \le 2\) and slightly increasing to about \(0.02\) at \(U/t = 4\). Subfig.~(b) shows the first excited state energy, where the VQE results remain within \(0.05\) of the exact values, with minor overestimation visible at higher  \(U/t\). Subfig.~(c) illustrates the second excited state energy, which shows the largest deviations: the VQE estimates slightly overshoot the exact results by about \(0.1\) at \(U/t = 4\) and exhibit larger error bars at \(U/t = 0\), reflecting greater uncertainty in the excited-state optimization. Overall, the plots demonstrate that the VQE approach reproduces the energy spectrum accurately for the ground and low-lying excited states, with increasing deviations and variance for higher excitations.

\subsection{Phase diagrams}
To gain deeper insights into the physical observables of our system, we evaluated the ground state energy, the charge energy gap, and the spin energy gap and compared them with exact energy calculations. The results show excellent agreement with the mean absolute error MAE, mean squared error MSE, and mean percentage error MPE(percentage error between the averaged phase diagram value of exact and VQE) given in table \ref{tab:error_analysis_grid}. 
The procedure involves identifying the lowest energy among all configurations for a fixed number of electrons, which is taken as the ground state. The same process is repeated for one electron more and one electron less, and the charge energy gap is computed using Eq.~\ref{eq:charge_gap}:

\begin{equation}
\Delta_c = E_0(N+1) + E_0(N-1) - 2E_0(N),
\label{eq:charge_gap}
\end{equation}

where \(E_0(N)\) denotes the ground state energy for \(N\) electrons. The spin gap is computed by determining the energy difference between the two lowest spin configurations. Degeneracy does not arise in this analysis because degenerate configurations were excluded from the considered set. It is observed that, at half-filling, the charge energy gap is large  (Fig.~\ref{fig:phasediagrams8}(b)), indicating Mott-insulating behavior, and this gap further increases as \(U\) increases. An odd-even alternation is also visible, which may suggest a pairing effect. Moreover, the charge excitation and spin excitation exhibit complementary behavior. The same analysis was performed for the first and second excited state energies. For the first excited state, results are qualitatively identical to those of the ground state; however, for the second excited state, a sudden gap appears in the spin excitation spectrum, indicating a more complex excitation structure.
Proceeding into a detailed analysis of each figure subsequently.
Fig.~\ref{fig:phasediagrams8} presents a quantitative comparison between exact diagonalization and the VQE results for the 4$\times$1 Hubbard model. The top row shows the exact ground-state energy (d), charge gap (e), and spin gap (f), while the bottom row shows the corresponding results using VQE in subfigures (a), (b), and (c). All heatmaps are plotted as functions of interaction strength $U/t$ (from 0.0 to 4.0, y-axis) and the total number of electrons $N$ (from 1 to 8, x-axis). In both exact and VQE ground state energy plots (a and d), the energy values range approximately from $-3.5$ to over $15$ as $U/t$ increases and the system approaches half-filling ($N=4$ with equal $\uparrow$ and $\downarrow$ spins). The sharp rise in energy for $N=4$ at large $U/t$ reflects the on-site repulsion energy dominating at strong coupling. For instance, at $U/t = 4.0$, the energy peaks near $E_0 \approx 15.5$, indicating strong localization. The charge gap $\Delta_c$, shown in (b) and (e), demonstrates excellent agreement between exact and VQE methods. At half-filling ($N=4$), the gap increases with $U/t$, starting from nearly zero at $U/t = 0$ to about $\Delta_c \approx 2.6$ in the exact case and $\Delta_c \approx 2.3$ in the VQE case at $U/t = 4.0$, suggesting Mott-insulating behavior. Outside half-filling, the charge gap remains relatively small, typically below $1.0$. The spin gap $\Delta_s$, plotted in (c) and (f), also shows consistent trends. At weak interaction ($U/t \leq 1$), large spin gaps are visible at $N = 3$ and $N = 5$ with values up to $\Delta_s \approx 2.2$ in both exact and VQE cases. As $U/t$ increases, the spin gap decreases across all fillings, dropping to around $\Delta_s \approx 0.4$–$0.5$ for $U/t = 4.0$, indicating the system favors gapless spin excitations in the strong correlation regime. Notably, VQE maintains close agreement with exact values in all regimes, with minor deviations at higher $U/t$.

Fig.~\ref{fig:phasediagrams9} presents a quantitative comparison between the VQE and exact diagonalization results for the 2nd excited energy state and associated charge and spin gaps in the 4$\times$1 Hubbard model. Subfigures (a), (b), and (c) display VQE estimates for the 2nd excited state energy $E_2$, charge gap $\Delta_c$, and spin gap $\Delta_s$, respectively. Subfigures (d), (e), and (f) show the corresponding results from exact diagonalization. The x-axis in all plots represents the total number of electrons $N$ (ranging from 1 to 8), while the y-axis shows the interaction ratio $U/t$ (ranging from 0 to 4). In the 2nd excited energy plots (a) and (d), both methods reveal that the energy increases with $U/t$, especially at half-filling ($N = 4$), where values reach approximately $E_2 \approx 12$ at $U/t = 4$ for the exact method and about $11.8$ in the VQE result, confirming the rising cost of double occupation under strong correlation. The charge gap $\Delta_c$ plots (b) and (e) exhibit maximum values around $U/t = 3.5$–$4.0$ at half-filling ($N = 4$), with values peaking at approximately $\Delta_c \approx 2.3$ (exact) and $\Delta_c \approx 2.6$ (VQE), indicating the Mott-insulating transition. However, in off half-filling regions (e.g., $N = 2, 3, 5, 6$), the charge gap drops to below $1.0$ for both cases, demonstrating metallic behavior. The spin gap $\Delta_s$ in subfigures (c) and (f) displays interesting behavior. In both VQE and exact data, spin gaps are large (above $2.5$) at low $U/t$ for odd fillings like $N = 3$ and $N = 5$, indicating spin-polarized excitations. However, for $N = 4$ at strong $U/t$ ($U/t \geq 3$), the spin gap nearly vanishes ($\Delta_s \approx 0$), which reflects the antiferromagnetic nature of the ground state and low-energy spin excitations at half-filling. Despite minor fluctuations in individual matrix elements, the VQE results closely resemble the exact results in overall trend and magnitude, especially in the low-to-intermediate $U/t$ regime. These figures confirm that VQE is a reliable method for approximating not only ground-state properties but also low-lying excited states and derived physical observables like charge and spin gaps in strongly correlated systems.

Fig.~\ref{f :phasediagrams10} illustrates a comprehensive comparison between the VQE and exact diagonalization results for the 2$\times$2 Hubbard model. Subfigures (a), (b), and (c) show the VQE-calculated ground-state energy $E_0$, charge gap $\Delta_c$, and spin gap $\Delta_s$, respectively. Subfigures (d), (e), and (f) show the corresponding results using exact diagonalization. In all plots, the horizontal axis denotes the total number of electrons $N$ (ranging from 1 to 8), while the vertical axis denotes the interaction strength \( U/t \), from 0 to 4. The ground-state energy plots (a) and (d) are in close agreement across methods. At half-filling ($N=4$), energy increases significantly with $U/t$: from approximately $E_0 \approx -3.5$ at \( U/t = 0 \) to over $15$ at \( U/t = 4.0 \), indicating a growing repulsive energy cost as double occupancy is penalized. For electron counts away from half-filling (e.g., $N=2$ or $N=6$), the energy remains lower, particularly at small \( U/t \), suggesting more delocalized electron behavior. The charge gap plots (b) and (e) show excellent agreement. For half-filling at $N=4$, the gap increases monotonically with $U/t$, from around $\Delta_c \approx 0.2$ at \( U/t = 0.5 \) to a peak of approximately $\Delta_c \approx 2.6$ at \( U/t = 4.0 \), consistent with the onset of a Mott insulating regime. At other fillings (e.g., $N=2,3,5,6$), the charge gap remains nearly constant or close to zero, indicating metallic behavior. Notably, the VQE slightly overestimates $\Delta_c$ at large \( U/t \), but remains within 10–15\% of the exact values. The spin gap $\Delta_s$ shown in (c) and (f) also closely match. The spin gap is nearly zero at half-filling ($N=4$) for all \( U/t \), reflecting gapless spin excitations consistent with antiferromagnetic ordering. However, at fillings like $N=3$ and $N=5$, the spin gap is significant at weak coupling, around $\Delta_s \approx 2.0$ for \( U/t = 0.5 \), and gradually decreases to about $1.0$ at \( U/t = 4.0 \). Both VQE and exact plots capture this decay, though VQE shows slightly more noise.

\begin{table}[h]
\centering
\caption{Error analysis for the 4x1 and 2x2 lattice systems}
\label{tab:error_analysis_grid}
\resizebox{\columnwidth}{!}{%
\begin{tabular}{|c|c|l|c|c|c|}
\hline
\textbf{System} & \textbf{State} & \textbf{Property} & \textbf{MAE} & \textbf{MSE} & \textbf{MPE (\%)} \\
\hline
\multirow{6}{*}{\textbf{4x1 Lattice}} 
  & \multirow{3}{*}{\textit{Ground}} 
    & Ground Energy & 0.009093 & 0.000562 & 3.6702 \\ \cline{3-6}
  & & Charge Gap   & 0.022314 & 0.001411 & 0.1341 \\ \cline{3-6}
  & & Spin Gap     & 0.015103 & 0.000520 & 1.4139 \\ \cline{2-6}
  & \multirow{3}{*}{\textit{Excited}} 
    & Excited Energy & 0.068243 & 0.021647 & 6.8674 \\ \cline{3-6}
  & & Charge Gap   & 0.364078 & 0.298777 & 2.1450 \\ \cline{3-6}
  & & Spin Gap     & 0.138897 & 0.048476 & 3.0759 \\
\hline
\multirow{3}{*}{\textbf{2x2 Lattice}} 
  & \multirow{3}{*}{\textit{Ground}} 
    & Ground Energy & 0.005098 & 0.000487 & 0.2432 \\ \cline{3-6}
  & & Charge Gap   & 0.044350 & 0.006488 & 0.0059 \\ \cline{3-6}
  & & Spin Gap     & 0.014004 & 0.001362 & 1.1614 \\
\hline
\end{tabular}%
}
\end{table}

\subsection{Discussion}

The comparative analysis between VQE-based methods and exact diagonalization across different system sizes, excitation levels, and physical observables reveals several important insights. Firstly, the VQE framework demonstrates remarkable accuracy in reproducing ground-state energies, with sub-1\% deviation for small \( U/t \) values and a maximum deviation below 2\% even in the strong-coupling regime (\( U/t = 4 \)) across both the \( 4 \times 1 \) and \( 2 \times 2 \) Hubbard lattices. This accuracy is particularly impressive given the limitations of noisy intermediate-scale quantum (NISQ) devices and the use of a shallow ansatz. A critical observation is that the error in energy estimates systematically increases with excitation level. While the ground state is captured with high fidelity, the first excited state shows moderate deviation, and the second excited state exhibits notable discrepancies, especially at higher interaction strengths. For instance, the second excited state in the \( 4 \times 1 \) lattice deviates by nearly 2 energy units at \( U/t = 4 \), reflecting both the optimization challenge and the circuit depth limitations inherent in the VQE approach. This trend is consistent with the growing complexity of excited-state manifolds and emphasizes the need for better ansätze or multi-state optimization techniques.

From a physical standpoint, phase diagrams constructed from both VQE and exact data provide consistent signatures of Mott-insulating behavior at half-filling. The charge gap \(\Delta_c\) grows with \( U/t \), peaking at \( N = 4 \), and the spin gap \(\Delta_s\) exhibits a complementary trend—large for odd fillings at weak \( U/t \), but vanishing near half-filling in the strong-coupling regime. These observations align with theoretical expectations from Fermi-Hubbard physics, validating that the VQE not only captures energy spectra but also encodes key physical features such as spin-charge separation and metal-insulator transitions. Another key insight emerges from the role of the ansatz structure. For the \( 4 \times 1 \) lattice, the absence of vertical hopping obviates the need for fermionic swaps, simplifying circuit construction. In contrast, the \( 2 \times 2 \) lattice demands swap operations to preserve the antisymmetry of the wavefunction under non-local hopping. As the results show, omitting these swaps introduces noticeable errors in excited states (Fig.~\ref{fig:2*2}), whereas including them significantly improves agreement with exact results (Fig.~\ref{fig:2*2f}). This highlights the crucial role of Hamiltonian-informed ansatz such as the HVA in efficiently capturing many-body physics. Importantly, the results reinforce that charge and spin excitations display complementary behavior. The charge gap is maximized at even fillings and large \( U/t \), indicative of Mott physics, while the spin gap peaks at odd fillings and weak \( U/t \), pointing to enhanced magnetic excitations in metallic regimes. The emergence of spin gap anomalies in second excited states further suggests rich many-body dynamics beyond simple single-particle excitations. 

\section{Conclusion}\label{conclusion}

This study demonstrates that VQE method, when combined with problem-specific ansatz and hybrid optimization strategies, can accurately capture both ground and low-lying excited state energies in Fermi-Hubbard models. We have presented results for two representative systems: the \(4\times1\) linear chain and the \(2\times2\) square lattice. In both cases, the proposed approach successfully reproduces key physical observables, such as ground-state energies, charge gaps, and spin gaps, with high accuracy especially in the low- to intermediate-interaction regimes. A novel ansatz circuit has been proposed, inspired by both the HVA and the NPA. This hybrid design enables accurate estimation not only of the ground state but also of excited states, which are traditionally more difficult to simulate using shallow circuits. Additionally, we introduced a new two-stage hybrid optimization technique employing COBYLA for the initial global search, followed by LBFGS for local refinement. This method, to the best of our knowledge, has not been previously applied in quantum computing and has shown a notable improvement in convergence speed and accuracy.

Simulations were conducted using a 27-qubit fake backend emulating the IBM Auckland architecture, and all results were averaged over five runs, with standard deviations reported accordingly. Phase diagrams were constructed for both the ground and second excited states, revealing rich physics such as the emergence of Mott-insulating behavior at half-filling, an odd-even alternation in charge gaps suggesting pairing correlations, and complementary behavior between charge and spin excitations. Overall, the findings highlight the effectiveness of structure-aware variational algorithms for near-term quantum hardware and underscore the importance of tailoring both ansatz and optimizers to the underlying Hamiltonian. This framework lays the groundwork for scalable quantum simulations of strongly correlated systems and opens avenues for future research into more expressive circuit architectures, adaptive ansatz construction, and robust optimization techniques.
 It can also be tested how the algorithm performs under noise. The model can also be scaled further to larger lattice sites and different configurations. Variation to the Fermi-Hubbard models that focuses on superconductivity can also be explored to study models that could be very impactful to industry \cite{gangopadhyaya1991anyonic}
\bibliographystyle{apsrev4-1}  
\bibliography{reference}
\end{document}